\documentclass[12pt,a4paper]{article}

\usepackage{graphicx}
\usepackage{amsmath}
\usepackage{amssymb}
\usepackage{cite}
\usepackage{hyperref}
\usepackage[height=8.8in,width=6.45in]{geometry}

\numberwithin{equation}{section}

\def\inc#1{\vcenter{\hbox{\includegraphics[scale=.25]{#1}}}}

\def\diag{\mathop{\mathrm{diag}}\nolimits}
\def\tr{\mathop{\mathrm{tr}}\nolimits}
\def\rank{\mathop{\mathrm{rank}}\nolimits}
\def\CP{\mathbb{CP}}
\def\GL{\mathrm{GL}}
\def\SL{\mathrm{SL}}
\def\SU{\mathrm{SU}}
\def\SO{\mathrm{SO}}
\def\USp{\mathrm{USp}}
\def\UU{\mathrm{U}}

\def\bC{\mathbb{C}}
\def\bR{\mathbb{R}}
\def\bZ{\mathbb{Z}}
\def\cH{\mathcal{H}}
\def\cS{\mathcal{S}}
\def\cM{\mathcal{M}}
\def\cDI{\mathcal{DI}}
\def\cN{\mathcal{N}}
\def\cV{\mathbb{V}}
\def\fk{\mathfrak{k}}
\def\fg{\mathfrak{g}}
\def\fh{\mathfrak{h}}
\def\sN{\text{\sffamily N}}
\def\bra#1{\langle#1|}
\def\ket#1{|#1\rangle}
\def\vev#1{\langle#1\rangle}
\def\Hyp{\mathcal{V}}
\def\ii{\mathrm{i}}

\begin{document}
\begin{titlepage}

\begin{flushright}
IPMU-14-0030\\
UT-14-6\\
\end{flushright}

\vskip 3cm

\begin{center}
{\Large \bfseries
A review on instanton counting and W-algebras
}

\vskip 1.2cm

Yuji Tachikawa$^{\sharp,\flat}$

\bigskip
\bigskip

\begin{tabular}{ll}
$^\flat$  & Department of Physics, Faculty of Science, \\
& University of Tokyo,  Bunkyo-ku, Tokyo 133-0022, Japan\\
$^\sharp$  & Kavli Institute for the Physics and Mathematics of the Universe, \\
& University of Tokyo,  Kashiwa, Chiba 277-8583, Japan
\end{tabular}
\vskip 1.5cm

\textbf{abstract}
\end{center}

\medskip
\noindent

Basics of the instanton counting and its relation to W-algebras are reviewed, with an emphasis toward physics ideas.
We discuss the case of $\UU(N)$ gauge group on $\bR^4$ to some detail, and indicate how it can be generalized to other gauge groups and to other spaces. 

This is part of a combined review on the recent developments on exact results on $\cN{=}2$ supersymmetric gauge theories, edited by J.~Teschner.

\bigskip
\vfill
\end{titlepage}

\setcounter{tocdepth}{2}

\tableofcontents
\section{Introduction}
\subsection{Instanton partition function}
After the indirect determination of the low-energy prepotential  of $\cN=2$ supersymmetric $\SU(2)$ gauge theory in \cite{Seiberg:1994rs,Seiberg:1994aj}, countless efforts were spent in obtaining the same prepotential in a much more direct manner, by performing the path integral over instanton contributions. After the first success in the 1-instanton sector \cite{Finnell:1995dr,Ito:1996qj}, people started developing techniques to perform  multi-instanton computations.  Years of study culminated in the publication of the review \cite{Dorey:2002ik} carefully describing both the explicit coordinates of and the integrand on the multi-instanton moduli space.  

A parallel development was ongoing around the same time,
which utilizes a powerful mathematical technique, called equivariant localization, in the instanton calculation.
 In \cite{Moore:1997dj}, the authors studied equivariant integrals over various hyperk\"ahler manifolds, including the instanton moduli spaces. From the start, their approach utilized the equivariant localization, but it was not quite clear at that time exactly which physical quantity they computed. Later in \cite{Losev:1997tp,Lossev:1997bz,Losev:1999nt}, the relation between the localization computation and the low-energy Seiberg-Witten theory was explored. Finally, there appeared the seminal paper by Nekrasov \cite{Nekrasov:2002qd}, where it was pointed out that the equivariant integral in \cite{Moore:1997dj}, applied to the instanton moduli spaces, is exactly the integral in \cite{Dorey:2002ik} which can be used to obtain the low-energy prepotential.  

In \cite{Nekrasov:2002qd}, a physical framework was also presented, where the appearance of the equivariant integral can be naturally understood. Namely, one can deform the theory on $\bR^4$ by two parameters $\epsilon_{1,2}$, such that a finite partition function $Z(\epsilon_{1,2};a_i)$ is well-defined, where $a_i$ are the special coordinates on the Coulomb branch of the theory. Then, one has \begin{equation}
\log Z(\epsilon_{1,2};a_i) \to \frac{1}{\epsilon_1\epsilon_2} F(a_i) + \text{less singular terms} \label{instanton-fundamental-relation}
\end{equation} in the $\epsilon_{1,2}\to 0$ limit. The function $Z(\epsilon_{1,2})$ is called under various names, such as \emph{Nekrasov's partition function}, the \emph{deformed partition function}, or the \emph{instanton partition function}. 
As the partition function is expressed as a discrete, infinite sum over instanton configurations, the method is dubbed \emph{instanton counting}.
In \cite{Flume:2001kb,Hollowood:2002ds,Flume:2002az,Hollowood:2002zv}, it was also noticed that the integral presented in \cite{Dorey:2002ik} is the integral of an equivariant Euler class, but the crucial idea of using $\epsilon_{1,2}$ is due to \cite{Nekrasov:2002qd}. 

For $\SU(N)$ gauge theory with fundamental hypermultiplets, the function $Z$ can be explicitly written down \cite{Nekrasov:2002qd,Flume:2002az,Bruzzo:2002xf,Nakajima:2003uh}.
The equality of the prepotential as defined by \eqref{instanton-fundamental-relation} and the prepotential as determined by the Seiberg-Witten curve is a rigorous mathematical statement which was soon proven by three groups by three distinct methods \cite{Nakajima:2003pg,Nekrasov:2003rj,Braverman:2004vv,Braverman:2004cr}. 
The calculational methods were soon generalized to quiver gauge theories, other matter contents, and other classical gauge groups \cite{Fucito:2004gi,Marino:2004cn,Nekrasov:2004vw,Shadchin:2004yx,Shadchin:2005cc,Hollands:2010xa,Hollands:2011zc}. It was also extended to calculations on the orbifolds of $\bR^4$ in \cite{Fucito:2004ry}. We now also know  a uniform derivation of the Seiberg-Witten curves from the instanton counting for $\SU$ quiver gauge theories with arbitrary shape  thanks to \cite{Fucito:2012xc,Nekrasov:2012xe}. Previous summaries and lecture notes on this topic can be found e.g.~in \cite{Shadchin:2005hp,Bianchi:2007ft}.

An $\cN=2$ gauge theory can often be engineered by considering  type IIA string on an open Calabi-Yau. It turned out \cite{Iqbal:2003ix,Iqbal:2003zz,Eguchi:2003sj,Iqbal:2004ne} that the topological A-model partition function as calculated by the topological vertex \cite{Iqbal:2002we,Aganagic:2003db} is then equal to Nekrasov's partition function of the five-dimensional version of the theory, when $\epsilon_1=-\epsilon_2$ is identified with the string coupling constant in the A-model. This suggested the existence of a refined, i.e.~two-parameter version of the topological string, and a refined formula for the topological vertex was formulated in \cite{Awata:2005fa,Iqbal:2007ii,Taki:2007dh,Awata:2008ed,Awata:2011ce}, so that the refined topological A-model partition function equals Nekrasov's partition function at $\epsilon_1+\epsilon_2\neq 0$. The relation between instanton partition functions and refined topological vertex was further studied in e.g.~\cite{Kozcaz:2010af,Taki:2010bj}. The same quantity can be computed in the mirror B-model side using the holomorphic anomaly equation \cite{Grimm:2007tm,Huang:2009md,Huang:2010kf,Huang:2011qx}, which also provided an independent insight to the system. 

We will derive the instanton partition function of four-dimensional gauge theories by  considering a five-dimensional system and then taking the four-dimensional limit.  
Therefore the review should prepare the reader so that they can understand systems in either dimensions. 
In this review, we mostly concentrate on four-dimensional theories, with only a cursory mention of the systems in five dimensions. 

\subsection{Relation to W-algebras}
Another recent developments concerns the two-dimensional CFT structure on the instanton partition function, which was first observed in \cite{Alday:2009aq,Gaiotto:2009ma} in the case of $\SU(2)$ gauge theory on $\bR^4$, and soon generalized to $\SU(N)$ in \cite{Wyllard:2009hg}, to other classical groups by \cite{Hollands:2010xa,Hollands:2011zc}, and to arbitrary gauge groups by \cite{Keller:2011ek}. 

This observation  was motivated from a general construction found in \cite{Gaiotto:2009we} and reviewed in \cite{G13,Ne} in this volume. Namely,  the 6d $\cN{=}(2,0)$ theory compactified on a Riemann surface $C$ gives rise to 4d $\cN{=}2$ theories labeled by $C$. Put the 4d theories thus obtained on $S^4$. The partition function can be computed as described in \cite{Pestun:2007rz,Hama:2012bg} and reviewed in \cite{PV}, which is given by an integral of the one-loop part and the instanton part. The one-loop part is given by a product of double-Gamma functions, and the instanton part is the product (one for the north pole and the other for the south pole) of two copies of the instanton partition function as reviewed in this review. As the one-loop part happens to be equal to that of the Liouville-Toda conformal field theory on $C$ as is reviewed in \cite{T}, the instanton part should necessarily be equal to the conformal blocks of these CFTs.
The conformal blocks have a strong connection to matrix models, and therefore the instanton partition functions can also be analyzed from this point of view. This will be further discussed in \cite{M} in this volume. 

We can also consider instanton partition functions of gauge group $\UU(N)$ on $\bR^4/\bZ_n$ where $\bZ_n$ is an subgroup of $\SU(2)$ acting on $\bR^4\simeq \bC^2$. 
Then the algebra which acts on the moduli space is guessed to be the so-called $n$-th para-$W_N$ algebra
 \cite{Belavin:2011pp,Bonelli:2011jx,Belavin:2011tb,Bonelli:2011kv,Wyllard:2011mn,Ito:2011mw,Alfimov:2011ju}.  
 For $\UU(2)$ on $\bR^4/\bZ_2$, we have definite confirmation that there is the action of a free boson, the affine algebra $\SU(2)_2$, together with the $\cN=1$ supersymmetric Virasoro algebra \cite{Belavin:2011pp,Belavin:2012aa}.

A further variation of the theme is to consider singularities in the  configuration of the gauge field along $\bC\subset \bC^2$. This is called a surface operator, and more will be discussed in \cite{G} in this volume. The simplest of these is characterized by the singular behavior $A_\theta d\theta \to \mu d\theta$ where $\theta$ is the angular coordinate transverse to the surface $\bC$ and $\mu$ is an element of the Lie algebra of the gauge group $G$ .  
The algebra which acts on the moduli space of instanton with this singularity is believed to be obtained by the Drinfeld-Sokolov reduction of the affine algebra of type $G$ \cite{Braverman:2010ef,Wyllard:2010vi,Kanno:2011fw}.  
In particular, when $\mu$ is a generic semisimple element, the Drinfeld-Sokolov reduction does not do anything in this case, and the algebra is the affine algebra of type $G$ itself when $G$ is simply-laced. This action of the affine algebra was constructed almost ten years ago \cite{Braverman:2004vv,Braverman:2004cr}, which was introduced to physics community in \cite{Alday:2010vg}.

\subsection*{Organization} 
We begin by recalling why the instantons configurations are important in gauge theory in Sec.~\ref{verybasic}. A rough introduction to the structure of the instanton moduli space is also given there. 
In Sec.~\ref{basics}, we  study the $\UU(N)$ gauge theory on $\bR^4$. 
We start in Sec.~\ref{toy} by considering the partition function of generic supersymmetric quantum mechanics.
In Sec.~\ref{first5d}, we will see how the instanton partition function reduces to the calculation of a supersymmetric quantum mechanics in general, which is then specialized to $\UU(N)$ gauge theory in Sec.~\ref{unitary}, for which explicit calculation is possible. The result is given a mathematical reformulation in Sec.~\ref{reformulation} in terms of the equivariant cohomology,
which is then given a physical interpretation in Sec.~\ref{physre}. The relation to the W-algebra is discussed in Sec.~\ref{WN}.
Its relation to the topological vertex is briefly explained in Sec.~\ref{top}; more details will be given in \cite{A} in this volume. 
In Sec.~\ref{othergauge} and Sec.~\ref{otherspace}, we indicate how the analysis can be extended to other gauge groups and to other spacetime geometries, respectively. 

Along the way, we will be able to see the ideas of three distinct mathematical proofs \cite{Nakajima:2003pg,Nekrasov:2003rj,Braverman:2004vv,Braverman:2004cr} of the agreement of the prepotential as obtained from the instanton counting and that as obtained from the  Seiberg-Witten curve. The proof by Nekrasov and Okounkov will be indicated in Sec.~\ref{unitary},
the proof by Braverman and Etingof in Sec.~\ref{surface}, and the proof by Nakajima and Yoshioka in Sec.~\ref{toric}.

In this paper we are not going to review standard results in W-algebras, which can all be found in \cite{Bouwknegt:1992wg,Bouwknegt:1995ag}.  The imaginary unit $\sqrt{-1}$ is denoted by $\ii$, as we will often use $i$ for the indices to sum over. 

If the reader understands Japanese, an even more introductory account of the whole story can be found in \cite{Tachikawa:2011rp}. 

\section{Gauge theory and the instanton moduli space}\label{verybasic}

\subsection{Instanton moduli space}\label{instantonmoduli}
Let us first briefly recall why we care about the instanton moduli space. 
We are interested in the Yang-Mills theory with gauge group $G$, whose partition function is given by \begin{equation}
Z=\int [DA_\mu] e^{-S} \quad
\text{where}
\quad
S=\frac{1}{2 g^2} \int \tr F_{\mu\nu} F_{\mu\nu},\label{pathintegral}
\end{equation} or its supersymmetric generalizations.  
Configurations with smaller action $S$ contribute more significantly to the partition function.
Therefore it is important to find the action-minimizing configuration:
\begin{equation}
\tr F_{\mu\nu}F_{\mu\nu} = \frac12 \tr(F_{\mu\nu}\pm \tilde F_{\mu\nu})^2 \mp  \tr F_{\mu\nu} \tilde F_{\mu\nu} 
\ge \mp \tr F_{\mu\nu}\tilde F_{\mu\nu}. 
\end{equation} 
For a finite-action configuration, it is known that the quantity \begin{equation}
n:= -\frac1{16\pi^2} \int d^4 x \tr F_{\mu\nu} \tilde F_{\mu\nu} 
\end{equation} is always an integer for the standard choice of the trace $\tr$ for $\SU(N)$ gauge field. For other gauge groups, we normalize the trace symbol $\tr$ so that this property holds true.
Then we find
\begin{equation}
\int d^4 x \tr F_{\mu\nu} F_{\mu\nu} \ge 16\pi^2 |n| \label{bound}
\end{equation} which is saturated only when \begin{equation}
F_{\mu\nu} + \tilde F_{\mu\nu}=0 \quad
\text{or}
\quad F_{\mu\nu} - \tilde F_{\mu\nu}=0 
\label{SD}
\end{equation} depending if $n>0$ or $n<0$, respectively.
This is the instanton equation. As it sets the (anti-)self-dual part of the Yang-Mills field strength to be zero, it is also called the (anti)-self dual equation, or the (A)SD equation for short.

The equation is invariant under the gauge transformation $g(x)$. 
We identify two solutions which are related by gauge transformations such that $g(x)\to 1$ at infinity.
The parameter space of instanton solutions is called the instanton moduli space, and we denote it by $M_{G,n}$ in this paper. 

For the simplest case  $G=\SU(2)$ and $n=1$, a solution is parameterized by eight parameters, namely \begin{itemize}
\item four parameters for the center, parameterizing $\bR^4$,
\item one parameter for the size, parameterizing $\bR_{>0}$,
\item and three parameters for the global gauge direction $\SU(2)/\bZ_2\sim S^3/\bZ_2$. 
\end{itemize} The last identification by $\bZ_2$ is due to the fact that the Yang-Mills field is in the triplet representation and therefore the element $\diag(-1,-1)\in \SU(2)$  doesn't act on it. 
The instanton moduli space is then \begin{equation}
M_{\SU(2),1}= \bR^4 \times \bR^4/\bZ_2\label{21}
\end{equation} where we combined $\bR_{>0}$ and $S^3$ to form an $\bR^4$. 

As the equation \eqref{SD} is scale invariant, an instanton can be shrunk to a point. 
This is called the small instanton singularity, which manifests in \eqref{21} as the $\bZ_2$ orbifold singularity at the origin.

For a general gauge group $G$ and still with $n=1$, it is known that every instanton solution is given by picking an $\SU(2)$ 1-instanton solution and regarding it as an instanton solution of gauge group $G$ by choosing an embedding $\SU(2)\to G$. It is known that such embeddings have $4h^{\vee}(G)-5$ parameters, where $h^\vee(G)$ is the dual Coxeter number of $G$. Together with the position of the center and the size, we have $4h^\vee(G)$ parameters in total. Equivalently, the instanton moduli space $M_{G,1}$ is real $4h^\vee(G)$ dimensional. It is a product of $\bR^4$ and the minimal nilpotent orbit of $\fg_\bC$: this fact will be useful in Sec.~\ref{exceptional}.

When $n>0$, one way to construct such a solution is to take $n$ 1-instanton solutions with well-separated centers, superimpose them, and add corrections to satisfy the equation \eqref{SD} necessary due to its nonlinearity. It is a remarkable fact that this operation is possible even when the centers are close to each other. The instanton moduli space $M_{G,n}$ then has real $4h^\vee(G)n$ dimensions. 
There is a subregion of the moduli space where one out of $n$ instantons shrink to zero size, and gives rise to the small instanton singularity. There,  the gauge configuration  is given by a smooth $(n-1)$-instanton solution with a pointlike instanton put on top of it. Therefore, the small instanton singularity has the form \cite{Uhlenbeck} \begin{equation}
\bR^4\times M_{G,n-1} \subset M_{G,n}.
\end{equation}

\subsection{Path integral around instanton configurations}
Now let us come back to the evaluation of the path integral \eqref{pathintegral}.
We split a general gauge field $A_\mu$ of instanton number $n$ into  a sum\begin{equation}
A_\mu = A_\mu^\text{inst} + \delta A_\mu 
\end{equation} where $A_\mu^\text{inst}$ is the instanton solution closest to the given configuration $A_\mu$. When $\delta A_\mu$ is small, we have \begin{equation}
S = \frac{8\pi^2 |n|}{ g^2} + \int d^4 x[ \text{(terms quadratic in $\delta A_\mu$)} + \text{(higher terms)}]
\end{equation} and the path integral becomes \begin{equation}
Z=\int [DA_\mu] e^{-S} =\sum_n \int_{M_{G,n}} d^{4h^\vee(G)n}X \int [\delta A_\mu] e^{-\frac{8\pi^2  |n|}{g^2} + \cdots } 
\end{equation}where $X\in M_{G,n}$ labels an instanton configuration. 

It was 't Hooft who first tried to use this decomposition to study the dynamics of quantum Yang-Mills theory \cite{'tHooft:1976fv}. It turned out that the integral over the fluctuations $\delta A_\mu$ around the instanton configuration makes the computation in the strongly coupled, infrared region very hard in general. 

For a supersymmetric model with a weakly coupled region, however, the fermionic fluctuations and the gauge fluctuations cancel, and often the result can be written as an integral over $M_{G,n}$ of a tractable function with explicit expressions; the state of the art at the turn of the century was summarized in the reference \cite{Dorey:2002ik}.  
One place the relation between supersymmetry and the instanton equation \eqref{SD} manifests itself is the supersymmetry transformation law of the gaugino, which is roughly of the form \begin{equation}
\delta \lambda_\alpha = F_{\alpha\beta} \epsilon^\beta, \qquad
\delta \bar\lambda_{\dot\alpha} = F_{\dot\alpha\dot\beta} \bar\epsilon^{\dot\beta}.
\end{equation} Here, $F_{\alpha\beta}$ and $F_{\dot\alpha\dot\beta}$ are (A)SD components of the field strength written in the spinor notation. Therefore, if the gauge configuration satisfies \eqref{SD}, then depending on the sign of $n$, half of the supersymmetry corresponding to  $\epsilon^\alpha$ or $\epsilon^{\dot\alpha}$ remains unbroken. 
In general, in the computation of the partition function in a supersymmetric background, only configurations preserving at least some of the supersymmetry gives non-vanishing contributions in the path integral. This is the principle called the supersymmetric localization.  
In this review we approach this type of computation from a rather geometric point of view.

\section{$\UU(N)$ gauge group on $\bR^4$ }\label{basics}
\subsection{Toy models}\label{toy}
We will start by considering supersymmetric quantum mechanics, as we are going to reduce the field theory calculations to supersymmetric quantum mechanics on instanton moduli spaces in Sec.~\ref{first5d}.
\subsubsection{Supersymmetric quantum mechanics on $\bC^2$}\label{C2}
Let us first consider the quantum mechanics of a supersymmetric particle on $\bC^2$, parameterized by $(z,w)$. 
Let the supersymmetry be such that $z$, $w$ are invariant, and $(\bar z,\psi_{\bar z})$ and $(\bar w,\psi_{\bar w})$ are paired. 
This system also has global symmetries $J_1$ and $J_2$, such that $(J_1,J_2)=(1,0)$ for $z$ and $(J_1,J_2)=(0,1)$ for $w$.

Let us consider its supersymmetric partition function\begin{equation}
Z(\beta;\epsilon_1,\epsilon_2) = \tr_\cH (-1)^F e^{\ii \beta \epsilon_1 J_1} e^{\ii \beta \epsilon_2 J_2} 
\end{equation} where $\cH$ is the total Hilbert space. 
As there is a cancellation within the pairs $(\bar z,\psi_{\bar z})$ and $(\bar w,\psi_{\bar w})$, we have the equality \begin{equation}
Z(\beta;\epsilon_1,\epsilon_2) = \tr_{\cH_\text{susy}}  e^{\ii \beta \epsilon_1 J_1} e^{\ii \beta \epsilon_2 J_2} 
\end{equation} where $\cH_\text{susy}$ is the subspace consisting of supersymmetric states, which in this case is \begin{equation}
\cH_\text{susy} \simeq \bigoplus_{m,n\ge 0} \bC z^m w^n.\label{sHC2}
\end{equation} The partition function is then \begin{equation}
Z(\beta;\epsilon_1,\epsilon_2)=\frac{1}{1-e^{\ii \beta\epsilon_1}}\frac{1}{1-e^{\ii \beta\epsilon_2}}. \label{C2partition}
\end{equation} In the $\beta\to 0$ limit, we have \begin{equation}
(-\ii\beta)^2 Z(\beta;\epsilon_1,\epsilon_2) \to \frac{1}{\epsilon_1\epsilon_2}.\label{slC2}
\end{equation}

\subsubsection{Supersymmetric quantum mechanics on $\CP^1$}
Next, consider a charged supersymmetric particle moving on $S^2\simeq \CP^1$, under the influence of a magnetic flux of charge $j=0,\frac12,1,$ etc. Let us use the complex coordinate $z$ so that $z=0$ is the north pole and $z=\infty$ is the south pole. 
The supersymmetric Hilbert space is then \begin{equation}
\cH_\text{susy} \simeq \bigoplus_{k=0}^{2j} \mathbb{C} z^k (\partial_z)^{\otimes j},
\end{equation}
and is the spin $j$ representation of $\SU(2)$ acting on $\CP^1$. 
 Let the global symmetry $J$ to rotate $z$ with charge 1. Then we have \begin{equation}
Z(\beta;\epsilon) = \tr_{\cH_\text{susy}}  e^{\ii \beta \epsilon J}  = e^{\ii j\beta\epsilon}+e^{\ii (j-1)\beta\epsilon}+\cdots+e^{-\ii j\beta\epsilon}.
\end{equation} This partition function can be re-expressed as \begin{equation}
Z(\beta;\epsilon)= \frac{e^{+\ii j\beta\epsilon}}{1-e^{-\ii \beta\epsilon}} + \frac{e^{-\ii j\beta\epsilon}}{1-e^{+\ii \beta\epsilon}}.\label{CP1partition}
\end{equation} Its $\beta\to0$ limit is finite: \begin{equation}
Z(\beta;\epsilon)\to 2j+1.
\end{equation}

\subsubsection{Localization theorem}
These two examples illustrate the following \emph{localization theorem}: consider a quantum mechanics of a supersymmetric particle moving on a smooth complex space $M$ of complex dimension $d$ with isometry $\UU(1)^n$, under the influence of a magnetic flux corresponding to a line bundle $L$ on $M$. 
Then the space of the supersymmetric states is the space of holomorphic sections of $L$. When $L$ is trivial, it is just the space of holomorphic functions on $M$.

Assume the points fixed by $\UU(1)^n$ on $M$ are isolated. 
Denote the generators of $\UU(1)^n$ by $J_1,\ldots, J_n$. Then the following relation holds: \begin{equation}
Z(\beta;\epsilon_1,\ldots,\epsilon_n) \equiv \tr_{\cH} (-1)^F e^{\ii \beta\sum_i \epsilon_i J_i} 
= \sum_{p} \frac{e^{\ii \beta \sum_i j(p)_i\epsilon_i}}{\prod_{a=1}^d (1-e^{\ii \beta\sum_i k(p)_{i,a}\epsilon_i })},\label{localization}
\end{equation}  see e.g.~\cite{AtiyahBook}.
Here, the sum runs over the set of fixed points $p$ on $M$, and $j(p)_i$ and $k(p)_{i,a}$ are defined so that \begin{equation}
\tr_{TM|_p} e^{\ii \beta\sum_i \epsilon_i J_i} = \sum_{a=1}^d e^{\ii \beta\sum_i k(p)_{i,a}\epsilon_i }
\end{equation} and \begin{equation}
\tr_{L|_p} e^{\ii \beta\sum_i \epsilon_i J_i} = e^{\ii \beta\sum_i j(p)_i J_i} .
\end{equation}
 In the following, it is convenient to abuse the notation and identify a vector space and its character under $\UU(1)^{N}$. Then we can just write \begin{equation}
 TM|_p=\sum_{a=1}^d e^{\ii \beta\sum_i k(p)_{i,a}\epsilon_i },\qquad
 L|_p = e^{\ii \beta\sum_i j(p)_i J_i} .
\end{equation} We will also use $+$, $\times$, $-$ instead of $\oplus$, $\otimes$ and $\ominus$.

In \eqref{C2partition}, the only fixed point is at $(z,w)=(0,0)$, and in \eqref{CP1partition}, there are two fixed points, one at $z=0$ and $z=\infty$. It is easy to check that the general theorem reproduces \eqref{C2partition} and \eqref{CP1partition}.

It is also clear that in the $\beta\to 0$ limit, we have \begin{equation}
(-\ii\beta)^dZ(\beta;\epsilon_1,\ldots,\epsilon_n) \to  \sum_{p} \frac{1}{\prod_{a=1}^d \sum_i k(p)_{i,a}\epsilon_i },
\end{equation} which is zero if $M$ is compact. 

\subsubsection{Supersymmetric quantum mechanics on $\bC^2/\bZ_2$}\label{C2Z2}
Let us make the identification by the $\bZ_2$ action $(z,w)\sim (-z,-w)$ in the model of Sec.~\ref{C2}. Then the supersymmetric Hilbert space \eqref{sHC2} becomes \begin{equation}
\cH_\text{susy} =\bigoplus_{m,n:\ \text{even}}\bC z^m w^n \oplus \bigoplus_{m,n:\ \text{odd}}\bC z^m w^n
\end{equation} and the partition function is therefore \begin{equation}
Z(\beta;\epsilon_1,\epsilon_2)=\frac{1+e^{\ii \beta(\epsilon_1+\epsilon_2)}}{(1-e^{2\ii \beta\epsilon_1})(1-e^{2\ii\beta\epsilon_2})}.\label{C2Z2partition}
\end{equation} The $\beta\to 0$ limit is then \begin{equation}
(\ii\beta)^2 Z(\beta;\epsilon_1,\epsilon_2) \to \frac{1}{2\epsilon_1\epsilon_2}.\label{slC2Z2}
\end{equation} The additional factor 2 with respect to \eqref{slC2} is due to the $\bZ_2$ identification. 

The localization theorem is not directly applicable, as the fixed point $(z,w)=(0,0)$ is singular. Instead, take the blow-up $M$ of $\bC^2/\bZ_2$, which is the total space of the canonical line bundle of $\CP^1$. The space is now smooth, with two fixed points. At the north pole $n$, \begin{equation}
\tr_{TM|_n} e^{\ii \beta(\epsilon_1J_1+\epsilon_2J_2)} = e^{2\ii \beta\epsilon_1}+e^{-\ii\beta(\epsilon_1-\epsilon_2)},
\end{equation} and at the south pole $s$, \begin{equation}
\tr_{TM|_s} e^{\ii \beta(\epsilon_1J_1+\epsilon_2J_2)} = e^{\ii\beta(\epsilon_1-\epsilon_2)} + e^{2\ii \beta\epsilon_2}.
\end{equation} Then we have \begin{equation}
Z(\beta;\epsilon_1,\epsilon_2)=\frac{1}{(1-e^{2\ii \beta\epsilon_1})(1-e^{-\ii\beta(\epsilon_1-\epsilon_2)})}+
\frac{1}{(1-e^{\ii\beta(\epsilon_1-\epsilon_2)})(1-e^{2\ii \beta\epsilon_2})}
\end{equation} from the localization theorem, which agrees with \eqref{C2Z2partition}.

\subsection{Instanton partition function: generalities}\label{first5d}
\begin{figure}
\[
\inc{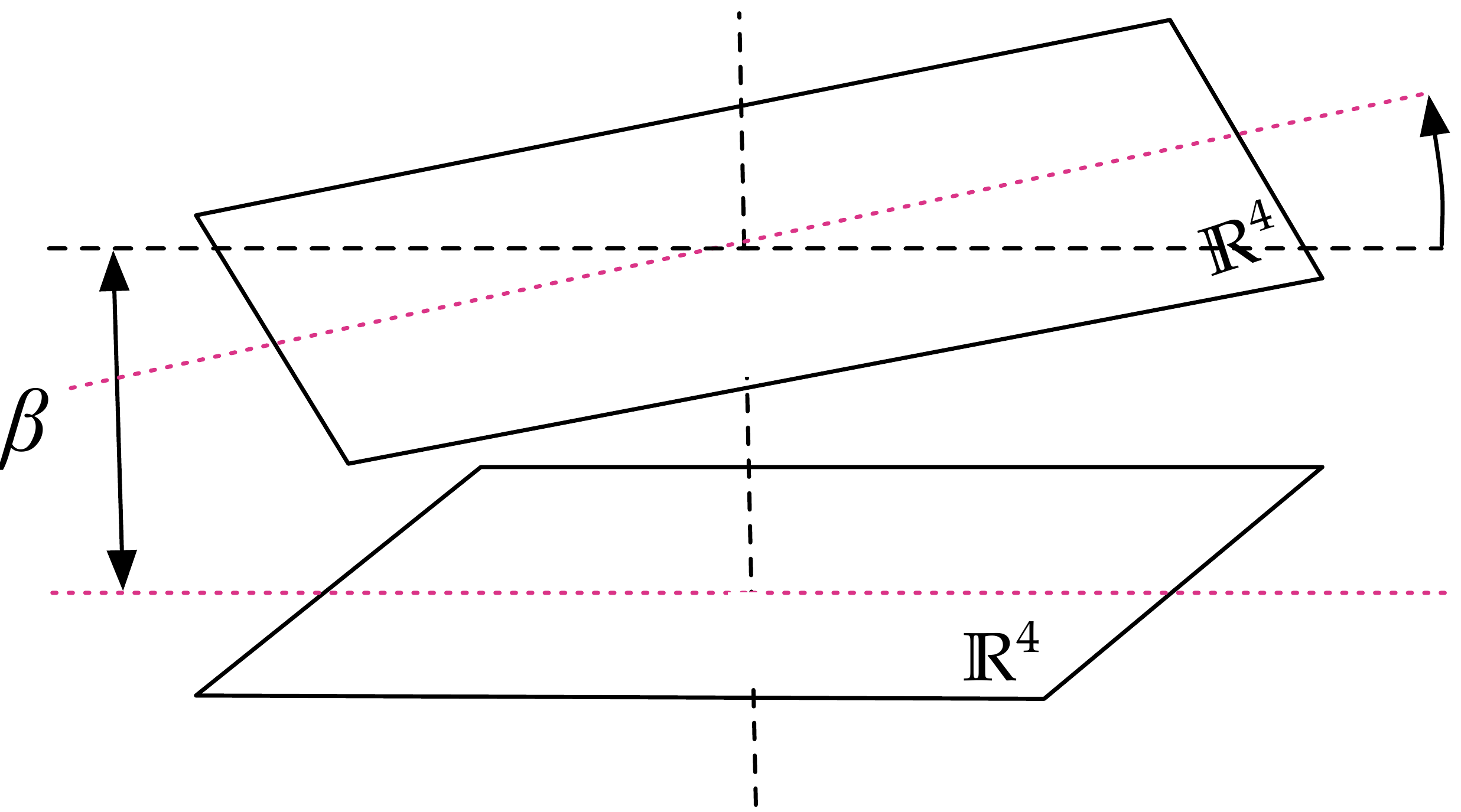}
\]
\caption{The five-dimensional spacetime. The vertical direction is $\xi^5$, and the $\bR^4$ planes at $\xi^5=0$ and $\xi^5=\beta$ are identified after a rotation.\label{omega}}
\end{figure}

Let us now come to the real objective of our study, namely the four-dimensional $\cN=2$ supersymmetric gauge theory.
The data defining the theory is its gauge group $G$, the flavor symmetry $F$, and the hypermultiplet  representation $R\oplus \bar R$ under $G\times F$.
With the same data, we can consider the five-dimensional $\cN=1$ supersymmetric gauge theory, with the same gauge group and the same hypermultiplet representation. 
We put this five-dimensional theory on a $\bC^2$ bundle over $S^1$ given by taking $\bC^2\times [0,\beta)$ parameterized by $(z,w,\xi^5)$, and making the identification \begin{equation}
(z,w,0)\sim (e^{\ii \beta\epsilon_1} z,e^{\ii \beta\epsilon_2} w, \beta).\label{omegaback}
\end{equation} See Fig.~\ref{omega} for a picture. This background space-time is often called the $\Omega$ background.

We set the vacuum expectation value of the gauge field at infinity, such that its integral along the $\xi^5$ direction is given by \begin{equation}
\diag(e^{\ii \beta a_1},e^{\ii \beta a_2},\ldots,e^{\ii \beta a_r}) \in \UU(1)^r \subset G.
\end{equation}  We also set the background vector field which couples to the flavor symmetry, such that its integral along the $\xi^5$ direction is given by \begin{equation}
\diag(e^{\ii \beta m_1},e^{\ii \beta m_2},\ldots,e^{\ii \beta m_f}) \in \UU(1)^f \subset F.
\end{equation} $m_i$ becomes the mass parameters when we take the four-dimensional limit $\beta\to 0$.

We are interested in the supersymmetric partition function in this background: \begin{equation}
Z(\beta;\epsilon_{1,2};a_{1,\ldots,r};m_{1,\ldots,f}) = \tr_{\cH_\text{QFT}} (-1)^F e^{\ii \beta(\epsilon_1J_1+\epsilon_2J_2+\sum_{s=1}^r a_s Q_s + \sum_{s=1}^f m_s F_s )}
\end{equation} where $\cH_\text{QFT}$ is the  Hilbert space of the five-dimensional field theory on $\bR^4$; $J_{1,2}$, $Q_{1,\ldots,r}$ and $F_{1,\ldots,f}$ are the generators of the spatial,  gauge and flavor rotation, respectively. 

We are mostly interested in the non-perturbative sector, where one has instanton configurations on $\bR^4$ with instanton number $n$.
Here  we assume that $G$ is a simple group; the generalization is obvious.

Energetically,  five-dimensional configurations which are close to a solution of the instanton equation \eqref{SD} at every constant time slice are favored within the path integral, similarly as discussed in Sec.~\ref{instantonmoduli}.  We can visualize such a configuration as one where the parameters describing the $n$-instanton configuration is slowly changing according to time. Therefore, the system can be approximated by the quantum mechanical particle moving within the instanton moduli space. 
This approach is often called the moduli space approximation.
With supersymmetry, this approximation becomes exact, and we have
\begin{equation}
Z_\text{inst}(\beta;\epsilon_{1,2};a_{1,\ldots,r})  
= \sum_{n\ge 0} e^{- \frac{8\pi^2 n\beta }{g^2}} \tr_{\cH_n} (-1)^F e^{\ii \beta(\epsilon_1J_1+\epsilon_2J_2+\sum_{s=1}^r a_s Q_s+ \sum_{s=1}^f m_s F_s )}
\end{equation} where $g$ is the five-dimensional coupling constant, and $\cH_n$ is the Hilbert space of the supersymmetric quantum mechanics on  the $n$-instanton moduli space.
Its  bosonic part $M_{G,n}$  is the moduli space of $n$-instantons of gauge group $G$, which we reviewed in Sec.~\ref{instantonmoduli}. It has  complex dimension $2h^\vee(G) n$.
In addition,  the fermionic direction $\Hyp(R)$ has complex dimension $k(R) n$, where $k(R)$ is the quadratic Casimir normalized so that it is $2h^\vee(G)$ for the adjoint representation. 
This $\Hyp(R)$ is a vector bundle over the instanton moduli space $M_{G,n}$, and is often called the matter bundle.

$M_{G,n}$ has a natural action of $\UU(1)^2$ which rotates the spacetime $\mathbb{C}^2$, and a natural action of $G$ which performs the spacetime independent gauge rotation. These actions extend equivariantly to the matter bundle $\Hyp(R)$.

Then, if $M_{G,n}$ were  smooth and if the fixed points $p$ under $U(1)^{2+r}\subset U(1)^2 \times G$ were isolated, we can apply the localization theorem to compute the instanton partition function: \begin{equation}
Z_\text{inst}(\beta;\epsilon_{1,2};a_{1,\ldots,r})  = \sum_{n\ge 0} e^{- \frac{8\pi^2 n\beta }{g^2}} 
\sum_p \frac{\prod_{t=1}^{k(R)n }  (1-e^{\ii \beta w_t(p) })}{\prod_{t=1}^{2h^\vee(G) n } (1-e^{\ii \beta v_t(p)})}
\label{5dpartition}
\end{equation} where $v_t(p)$ and $w_t(p)$  are linear combinations of $\epsilon_{1,2}$, $a_{1,\ldots,r}$ and $m_{1,\ldots,f}$ such that we have \begin{equation}
TM_{G,n}|_p= \sum_{t=1}^{2h^\vee(G) n}e^{\ii \beta v_t(p)}, \quad
\Hyp(R)|_p=\sum_{t=1}^{k(R) n}e^{\ii \beta w_t(p)}. \label{fpweights}
\end{equation}
As was explained in Sec.~\ref{instantonmoduli}, $M_{G,n}$ has small instanton singularities and the formula above is not directly applicable. One of the technical difficulties in the instanton computation is how to deal with this singularity.  Currently, the explicit formula is known (or, at least the method to write it down is known) for the following cases: i)  $G=\UU(N)$ with arbitrary representations, 
ii)  $G=\SO(N)$ with representations appearing in the tensor powers of the vector representation, and iii)
$G=\USp(2N)$ with arbitrary representations.   We will discuss $\UU(N)$ with (bi)fundamentals in Sec.~\ref{unitary}, and $\SO(N)$ and $\USp(2N)$ with fundamentals in Sec.~\ref{otherclassical}. For other representations, see \cite{Shadchin:2004yx,Shadchin:2005cc}.

The 5d gauge theory can have a Chern-Simons coupling, it induces a magnetic flux to the supersymmetric quantum mechanics on the instanton moduli space, which will introduce a factor in the numerator of \eqref{5dpartition} as dictated by the localization theorem \eqref{localization} \cite{Tachikawa:2004ur}.

The four-dimensional limit $\beta\to0$ needs to be taken carefully. 
In principle threre can be multiple interesting choices of the  scaling of the variables, resulting in different four dimensional dynamics. Here we only consider the standard one.
We would like to take the limit $\beta\to 0$ keeping $\epsilon_{1,2}$ and $a_i$ finite.
Note that each term in the sum \eqref{5dpartition} with fixed instanton number $n$ has $(2h^\vee(G)-k(R))n$ more factors in the denominator, producing a factor $\propto\beta^{-(2h^\vee(G)-k(R))n}$.
In order to compensate it, we express the classical contribution to the action in \eqref{5dpartition} as 
\begin{equation}
e^{- \frac{8\pi^2 \beta}{g^2}} = (-\ii\beta)^{2h^\vee(G)-k(R)} q \label{running}
\end{equation} and keep $q$ fixed while taking $\beta\to 0$.
The four-dimensional limit of the partition function is then 
\begin{equation}
Z_\text{inst}(\beta;\epsilon_{1,2};a_{1,\ldots,r})  = \sum_{n\ge 0} q^n
\sum_p \frac{\prod_{t=1}^{k(R)n }  w_t(p) }{\prod_{t=1}^{2h^\vee(G) n } v_t(p) }.\label{4dpartition}
\end{equation} 
 Note that the naive four-dimensional coupling $g_{4d}$ is given by the five-dimensional coupling  $g_{5d}$ by the relation \begin{equation}
 \frac{8\pi^2 }{g^2_\text{4d}} = \frac{8\pi^2 \beta}{g^2_\text{5d}}.
\end{equation} 
Therefore, the relation \eqref{running}, where $q$ is fixed and $\beta$ is varied, can be thought of as describing the running of $g_{4d}$ when we change the UV cutoff scale $\beta^{-1}$. 
We see that the relation \eqref{running} correctly reproduces the logarithmic one-loop running of $g_{4d}$, controlled by the one-loop beta function coefficient $2h^\vee(G)-k(R)$. 
The dynamical scale $\Lambda$ is given by $q=\Lambda^{2h^\vee(G)-k(R)}$.
It is somewhat gratifying to see that the logarithmic running arises naturally in this convoluted framework. 
 
This definition of the four-dimensional instanton partition function does not explain why its limit \begin{equation}
F(a_{1,\ldots,r}) = \lim_{\epsilon_{1,2}\to 0}\epsilon_1\epsilon_2 \log Z_\text{inst}(\epsilon_{1,2};a_{1,\ldots,r})
\end{equation} is the prepotential of the four-dimensional gauge theory.  For field theoretical explanations, see~\cite{Nekrasov:2002qd} or the Appendix of \cite{Tachikawa:2004ur}.

\subsection{Instanton partition function: unitary gauge groups }\label{unitary}
The instanton moduli space is always singular as explained in Sec.~\ref{instantonmoduli}.
Therefore, we need to do something in order to apply the idea outlined in the previous section.
When the gauge group is $\UU(N)$, there is a standard way to deform the singularities so that the resulting space is smooth \cite{Nekrasov:1998ss,Losev:1999nt}. 

\paragraph{ADHM construction} Let the instanton number be $n$, and introduce the space $M_{G,n,t}$ via \begin{equation}
M_{G,n,t} := \{\mu_\bC(x) = t \mid  x\in X_{G,n} \} / \GL(n).\label{ADHM}
\end{equation}
\begin{itemize}
\item Here $X_{G,n}$ is a linear space constructed from two vector spaces $V$, $W$ described below as follows \begin{equation}
X_{G,n}=(T_1^{\otimes-1}\oplus T_2^{\otimes-1})\otimes  V \otimes V^* \oplus  W^*\otimes   V\oplus T_1^{\otimes -1}\otimes T_2^{\otimes -1}  \otimes V^* \otimes W.
\end{equation} Here, $T_i$ is a one-dimensional space on which the generator $J_i$ has the eigenvalue $+1$. 
As it is very cumbersome to write a lot of $\otimes$ and $\oplus$, we abuse the notation as already introduced above, by identifying the vector space and its character:
\begin{equation}
X_{G,n}=(e^{-\ii\beta\epsilon_1}+e^{-\ii\beta\epsilon_2}) V V^* + W^*  V+ e^{-\ii\beta(\epsilon_1+\epsilon_2)} V^*  W\label{Xn}.
\end{equation}
\item $V\simeq \bC^n$ is a space with a natural $\GL(n)$ action and,
\item  $W\simeq \bC^N$ is a space with a natural $\UU(N)$ action.
\item The $*$ operation is defined naturally by setting $\ii^*=-\ii$, $\epsilon_{1,2}{}^*=\epsilon_{1,2}$, and $a_{1,\ldots,r}{}^*=a_{1,\ldots,r}$,
\item and $\mu_\bC$ is a certain quadratic function on $X_{G,n}$ taking value in the Lie algebra of $\GL(n)$,
\item and finally $t$ is a deformation parameter taking value in the center of the Lie algebra of $\GL(n)$. For generic $t$ the space is smooth, but it becomes  singular when $t=0$.
\end{itemize}
This is called the ADHM construction, and the space at $t=0$, $M_{G,n,0}$, is the instanton moduli space $M_{G,n}$. 

The trick we use is to replace $M_{G,n}$ by $M_{G,n,t}$ with $t\neq 0$ and apply the localization theorem. The answer does not depend on $t$ as long as it is non-zero. 
The deformation by $t$ can be physically realized by the introduction of the spacetime noncommutativity \cite{Nekrasov:1998ss}, but this physical interpretation does not play any role here.
Mathematically, this deformation corresponds to considering not just bundles but also torsion free sheaves, see e.g.~\cite{Losev:1999nt}.
Note that  it is not known how to perform such deformation in other gauge groups at present.

The fixed points of the  $\UU(1)^{2+N}$ action on $M_{G,n,t}$ was classified in \cite{Nakajima:2003uh}, which we will describe below. 
A fixed point $p$ is labeled by $N$ Young diagrams $\vec Y=(Y_1,\ldots,Y_N)$ such that the total number of the boxes $|\vec Y|$ is $n$. Let us denote by $(i,j)\in Y$ when there is a box at the position $(i,j)$ in a Young diagram $Y$. Then, the fixed point labeled by $p=(Y_1,\ldots,Y_N)$ corresponds to the action of $\UU(1)^2$ and $\UU(1)^r\subset G$ on $V$ and $W$ such that \begin{equation}
W_p= \sum_{s=1}^N e^{\ii \beta a_s }, \qquad
V_p = \sum_{s=1}^N \sum_{(i,j)\in Y_s} e^{\ii \beta (a_s +(1-i)\epsilon_1+(1-j)\epsilon_2)}.\label{WV}
\end{equation}

Then we have \begin{equation}
TM|_p =  W^*_p V_p  + e^{\ii \beta(\epsilon_1+\epsilon_2)}  V^*_p  W_p - (1-e^{\ii \beta\epsilon_1})(1-e^{\ii \beta\epsilon_2})  V_p V^*_p,\label{tangent}
\end{equation} from which you can read off $v(p)_t$ in \eqref{fpweights}. As for $w(p)_t$, we have \begin{equation}
\Hyp(\text{fundamental})_p= e^{-\ii\beta m}V_p,\qquad
\Hyp(\text{adjoint})_p= e^{-\ii\beta m}TM|_p\label{fa}
\end{equation} where $m$ is the mass of the hypermultiplets. 
In the case of a bifundamental of $\UU(N_1)\times \UU(N_2)$, the zero modes are determined once the instanton configurations $p,q$ of $\UU(N_{1,2})$ are specified: \begin{equation}
\Hyp(\text{bifundamental})_{p,q} = e^{-\ii\beta m}(
W^*_p V_q  + e^{\ii \beta(\epsilon_1+\epsilon_2)}  V^*_p  W_q  - (1-e^{\ii \beta\epsilon_1})(1-e^{\ii \beta\epsilon_2})  V_p^*  V_q).\label{bif}
\end{equation} 
Note that both the adjoint and the fundamental are  special cases of the bifundamental, namely, the adjoint is when $p=q$, and the fundamental is when $p$ is empty.

Then it is just a combinatorial exercise to write down the explicit formula for the four-dimensional partition function \eqref{4dpartition} in terms of Young diagrams labeling the fixed points. 
The explicit formulas are given below. 
However, before writing them down, the author would like to stress that to implement it in a computer algebra system, it is usually easier and less error-prone to just directly use the formulas \eqref{WV}, \eqref{tangent}, \eqref{fa}, \eqref{bif} to compute the characters and then to read off $v(p)_t$ and $w(p)_t$  via \eqref{fpweights}, which can then be plugged in to \eqref{4dpartition}. 

\begin{figure}
\centerline{\includegraphics[scale=.3]{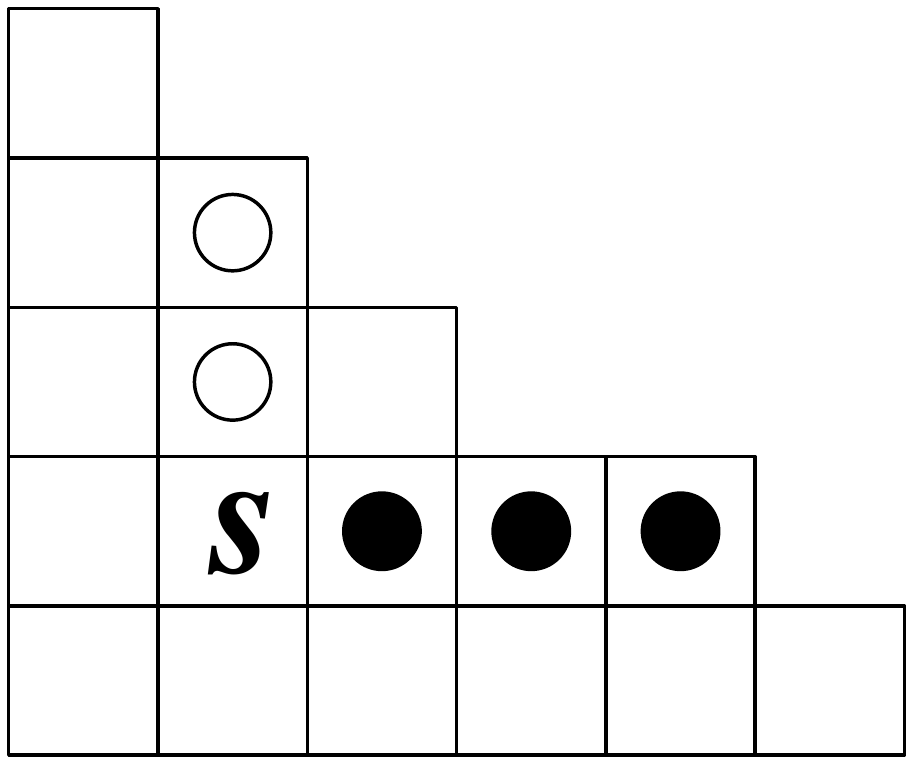}}
\caption{Definition of the arm-length and the leg-length. For a box $s$ in a Young tableau displayed above, the leg-length is the number of boxes to the right of $s$,
marked by black disks, and the arm-length is the number of boxes on top of $s$.\label{fig:hook}}
\end{figure}

\paragraph{Explicit formulas} 
Let $Y=(\lambda_1\ge \lambda_2\ge \cdots)$ be a Young tableau
where $\lambda_i$ is the height of the $i$-th column.
We set $\lambda_i=0$ when $i$ is larger than the width of the tableau.
Let $Y^T=(\lambda_1'\ge \lambda'_2\ge \cdots)$ be its transpose.
For a box $s$ at the coordinate $(i,j)$, we let its arm-length $A_Y(s)$
and leg-length $L_Y(s)$ with respect to the tableau $Y$ to be \begin{equation}
A_Y(s)=\lambda_i-j,\qquad
L_Y(s)=\lambda_j'-i,
\end{equation}  see Fig.~\ref{fig:hook}. Note that they can be negative when
the box $s$ is outside the tableau.
We then define a function $E$ by \begin{equation}
E(a,Y_1,Y_2,s)=a-\epsilon_1 L_{Y_2}(s) + \epsilon_2 (A_{Y_1}(s)+1).
\end{equation}
We use the vector symbol $\vec a$  to stand for $N$-tuples,
e.g.~ $\vec Y=(Y_1,Y_2,\ldots,Y_N)$, etc.

Then, the contribution of an $\SU(N)$ vector multiplet from the fixed point $p$ labeled by an $N$-tuple of Young diagrams $\vec Y$ is the denominator of \eqref{4dpartition}, where $v_t(p)$ can be read off from the characters of $TM_{G,n}|_{p}$ once we have the form \eqref{fpweights}. This is done by  plugging \eqref{WV} to \eqref{tangent}. 
The end result is 
 \begin{equation}
z_\text{vect}(\vec a,\vec Y)=\frac{1}{
\prod_{i,j=1}^N
\prod_{s\in Y_i}E(a_i-a_j,Y_i,Y_j,s)
\prod_{t\in Y_j}(\epsilon_1+\epsilon_2  -  E(a_j-a_i,Y_j,Y_i,t))
}.
\end{equation}  Note that there are $2Nn$ factors in total. This is as it should be, as $TM_{G,n}$ is complex $2Nn$ dimensional, and there are $2Nn$ eigenvalues at each fixed point.

The contribution from (anti)fundamental hypermultiplets is given by
\begin{align}
z_\text{fund}(\vec a,\vec Y,m)&=\prod_{i=1}^N \prod_{s\in Y_i} ( \phi(a_i,s) -m+\epsilon_1+\epsilon_2), \\
z_\text{antifund}(\vec a,\vec Y,m)&=
z_\text{fund}(\vec a,\vec Y,\epsilon_1+\epsilon_2-m)
\end{align} where $\phi(a,s)$ for the box $s=(i,j)$ is defined as \begin{equation}
\phi(a,s)=a+\epsilon_1 (i-1)+\epsilon_2(j-1).
\end{equation} They directly reflect the characters of $V_p$ in \eqref{WV}. 

When we have gauge group $\SU(N)\times \SU(M)$ and a bifundamental charged under both, the contribution from the bifundamental depends on the gauge configuration of both factors of the gauge group. 
Namely, for the fixed point $p$ of $M_{\SU(N),n,t}$ labeled by the Young diagram $\vec Y$
and 
the fixed point $q$ of $M_{\SU(N),n}$ labeled by the Young diagram $\vec W$,
the contribution of a bifundamental is \cite{Fucito:2004gi,Shadchin:2005cc}: \begin{multline}
z_\text{bifund}(\vec a,\vec Y;\vec b,\vec W;m)= \\
\prod_{i}^N \prod_j^M
\prod_{s\in Y_i}( E(a_i-b_j,Y_i,W_j,s)- m)
\prod_{t\in W_j}(\epsilon_1+\epsilon_2  -  E(b_j-a_i,W_j,Y_i,t)- m)
\end{multline} where $\vec a$ and $\vec b$ are the chemical potentials for $\SU(N)$ and $\SU(M)$ respectively.

The contribution of an adjoint hypermultiplet is a special case where $p=q$ and $\vec a=\vec b$.
It is \begin{equation}
z_\text{adj}(\vec a,\vec Y,m)=z_\text{bifund}(\vec a,\vec Y,\vec a,\vec Y,m).
\end{equation} This satisfies 
\begin{equation}
z_\text{vector}(\vec a,\vec Y)=1/z_\text{adj}(\vec a,\vec Y,0).
\end{equation}

Note that there are several definitions of the mass parameter $m$. Another definition with \begin{equation}
m'=m-\frac{1}{2}(\epsilon_1+\epsilon_2)
\end{equation} is also common. For their relative merits, the reader is referred to the thorough discussion in \cite{Okuda:2010ke}.

Let us write down, as an example, the instanton partition function of $\cN=2^*$ $\SU(N)$ gauge theory, i.e.~an $\SU(N)$ theory with a massive adjoint multiplet. We just have to multiply the contributions determined above, and we have \begin{multline}
Z=\sum_{n\ge 0} q^n \sum_{\vec Y, |\vec Y|=n} z_\text{adj}(\vec a,\vec Y,m) z_\text{vector}(\vec a,\vec Y) \\
= \sum_{\vec Y}  q^{|\vec Y|}
\prod_{i,j=1}^N\frac{
\prod_{s\in Y_i}(E(a_i-a_j,Y_i,Y_j,s)+m)
\prod_{t\in Y_j}(\epsilon_1+\epsilon_2  -  E(a_j-a_i,Y_j,Y_i,t)-m)
}{
\prod_{s\in Y_i}E(a_i-a_j,Y_i,Y_j,s)
\prod_{t\in Y_j}(\epsilon_1+\epsilon_2  -  E(a_j-a_i,Y_j,Y_i,t))
}.
\end{multline} 

\paragraph{Nekrasov-Okounkov} For $G=\UU(N)$, the final result is a summation over $N$-tuples of Young diagrams $p=(Y_1,\ldots, Y_N)$ of a rational function of $\epsilon_{1,2}$, $a_{1,\ldots,r}$ and $m_{1,\ldots,f}$. The prepotential can be extracted by taking the limit $\epsilon_{1,2}\to 0$. There, the summation can be replaced by an extremalization procedure over the asymptotic shape of the Young diagrams. Applying the matrix model technique, one finds that the prepotential as obtained from this instanton counting is the same as the prepotential as defined by the Seiberg-Witten curve \cite{Nekrasov:2003rj}.

\paragraph{Explicit evaluation for $\UU(2)$ with 1-instanton}
Before proceeding, let us calculate the instanton partition function for the pure $\UU(2)$ gauge theory at 1-instanton level explicitly. 
It would be a good exercise, as the machinery used so far has been rather heavy, and the formulas are although concrete rather complicated. 

In fact, the calculation is already done in Sec.~\ref{toy}, since the moduli space in question is $\bC^2\times \bC^2/\bZ_2$.
Here the first factor $\bC^2$ is the position of the center of the instanton, and $\bC^2/\bZ^2 \sim \bR_{>0} \times S^3/\bZ_2$ parameterizes the gauge orientation of the instanton via $S^3/\bZ_2\simeq \SO(3) \simeq \SU(2)/\bZ_2$ and the size of the instanton via $\bR_{>0}$. 
Introduce the coordinates $(z,w,u,v)$ with the identification $(u,v)\sim (-u,-v)$. 
The action of $e^{\ii \beta(\epsilon_1J_1+\epsilon_2J_2)}$  is given by \begin{equation}
(z,w,u,v)\to
(e^{\ii \beta\epsilon_1} z,e^{\ii \beta\epsilon_2} w,
e^{\ii \beta(\epsilon_1+\epsilon_2)/2} u,
e^{\ii \beta(\epsilon_1+\epsilon_2)/2} v)
\end{equation} and $(u,v)$ form a doublet under the $\SU(2)$ gauge group. Then for $\diag(e^{\ii \beta a},e^{-\ii\beta a})\in \SU(2)$,
we have \begin{equation}
(u,v)\to (e^{\ii \beta a}u,e^{-\ii\beta a}v).
\end{equation} Then the instanton partition function is given by combining \eqref{slC2} and \eqref{slC2Z2}: \begin{equation}
Z_\text{inst}(\epsilon_{1,2};a)=\frac{1}{\epsilon_1\epsilon_2}\frac12 \frac{1}{(\epsilon_1+\epsilon_2)/2-a}\frac{1}{(\epsilon_1+\epsilon_2)/2+a}.
\end{equation}
It is an instructive exercise to reproduce this from the general method explained earlier in this section. 

\subsection{A mathematical reformulation}\label{reformulation}
Let us now perform a mathematical reformulation, following the idea of \cite{CarlssonOkounkov}.
For $G=U(N)$, consider the vector space \begin{equation}
\cV_{G,\vec a}=\bigoplus_{n=0}^\infty \cV_{G,\vec a,n}
\end{equation} where \begin{equation}
\cV_{G,\vec a,n}=\bigoplus_{p} \bC\ket{p}
\end{equation} where $p$ runs over the fixed points of $\UU(1)^{2+r}$ action on $M_{G,n,t}$. 
We define the inner product by taking the denominator of \eqref{4dpartition}: \begin{equation}
\vev{p|q}=\delta_{p,q} \frac{1}{\prod_{t} v(p)_t}.\label{4dinner}
\end{equation} 
Note that the basis vectors are independent of $\vec a$, but the inner product does depend on $\vec a$.
We introduce an operator $\sN$ such that $\cV_{G,\vec a,n}$ is the eigenspace with eigenvalue $n$. 

Let us introduce a vector \begin{equation}
\ket{\text{pure}}=\sum_{n=0}^\infty \sum_{p} \ket{p} \in \cV_{G,\vec a}.
\end{equation} 

Then the partition function \eqref{4dpartition} of the pure $\SU(N)$ gauge theory is just \begin{equation}
Z(\epsilon_{1,2};\vec a)= \bra{\text{pure}}q^{\sN} \ket{\text{pure}}.\label{purepartition}
\end{equation}

A bifundamental charged under $G_1=\SU(N_1)$ and $G_2=\SU(N_2)$ defines a linear map
\begin{equation}
\Phi_{\vec b,m,\vec a}: \cV_{G_1,\vec a}\to \cV_{G_2,\vec b}\label{bifop}
\end{equation} such that \begin{equation}
\bra q\Phi_m\ket p = \prod_t w(p,q)_t
\end{equation} where the right hand side comes from the decomposition \begin{equation}
\Hyp(\text{bifundamental})_{p,q}=\sum_t e^{-\ii\beta w(p,q)_t}.
\end{equation}

\begin{figure}
\[
\begin{array}{rl}
\text{4d)} & \inc{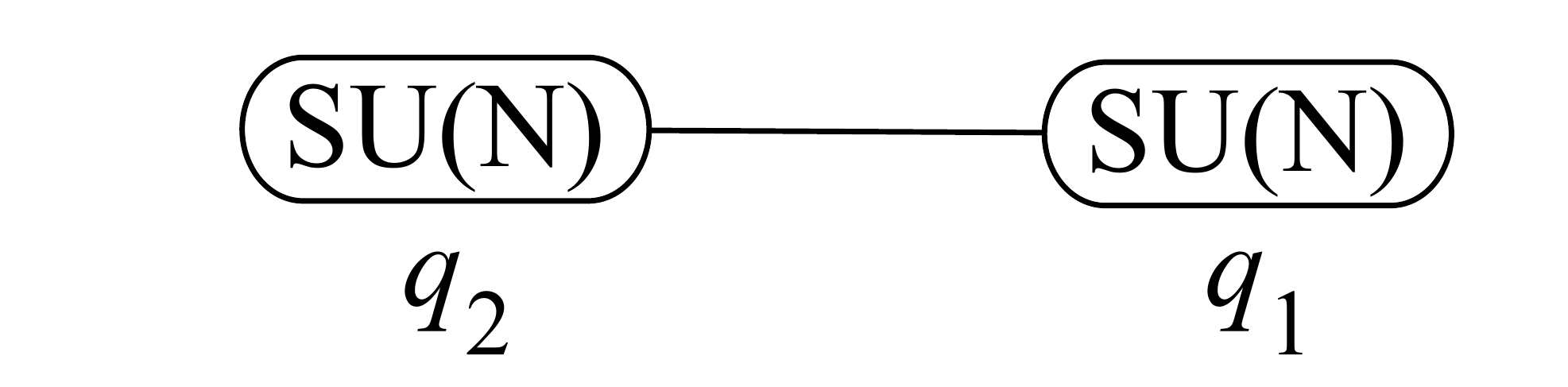}\\
\text{5d)} & \inc{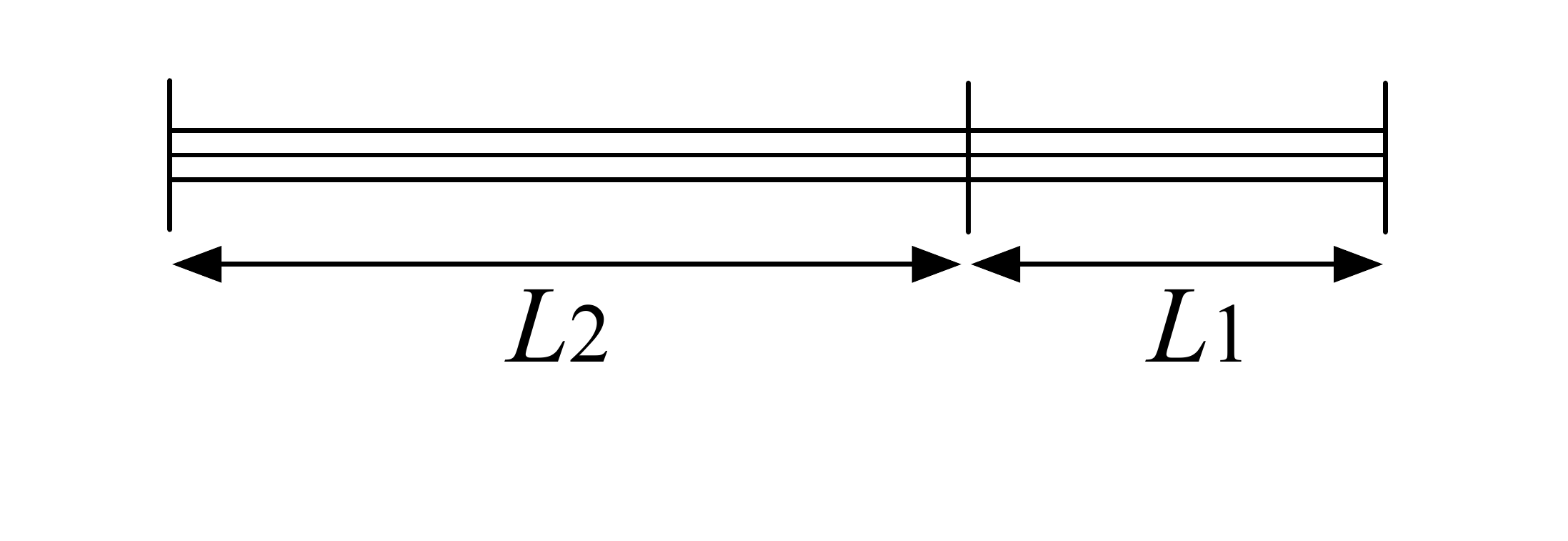}\\
\text{eq)} & \inc{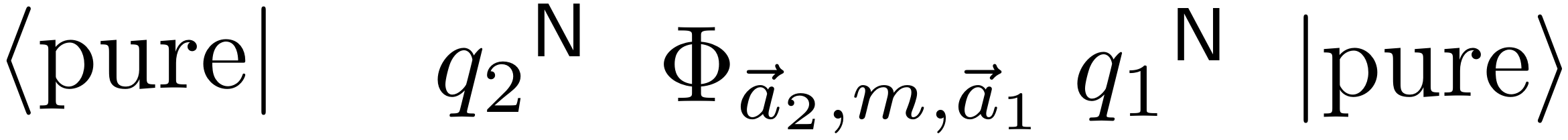}\\
\text{6d)}& \inc{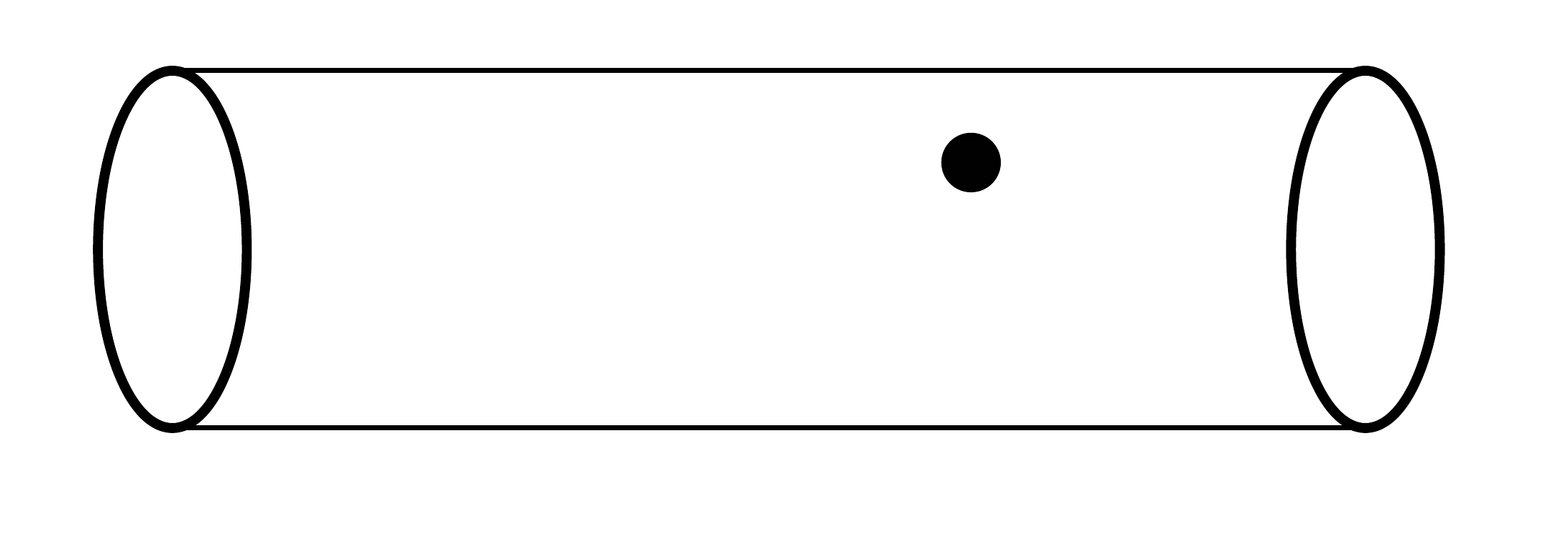}
\end{array}
\]
\caption{Higher-dimensional setup of the quiver theory.  The horizontal direction is $x^5$.  
4d) the $\SU(N)\times \SU(N)$ quiver gauge theory in the infrared.
5d) 5d maximally supersymmetric $\SU(N)$ theory on a segment.  eq) its partition function, regarding the fifth direction as ``time''. 6d) 6d $\cN=(2,0)$ theory on a cylinder.\label{higher}}
\end{figure}

Using this linear map $\Phi_{\vec b,m,\vec a}$, we can concisely express the partition function of  quiver gauge theories.  For example, consider $\SU(N)_1\times \SU(N)_2$ gauge theory with bifundamental hypermultiplets charged under $\SU(N)_1\times \SU(N)_2$ with mass $m$, see Fig.~\ref{higher} (4d).
Then the instanton partition function \eqref{4dpartition} is just \begin{equation}
Z(\epsilon_{1,2};\vec a_1,\vec a_2;\vec m)
=
\bra{\text{pure}}
q_2{}^{\sN}\Phi_{\vec a_2,m,\vec a_1} q_1{}^{\sN}
\ket{\text{pure}}.\label{quiverpartition}
\end{equation} 

In this section, we introduced the vector space $\cV_{G,\vec a,n}$ together with its inner product using fixed points of $M_{G,n,t}$. It is known that this vector space is a natural mathematical object called the equivariant cohomology:
\begin{equation}
\cV_{G,\vec a,n}=H^*_{G\times \UU(1)^2}(M_{G,n,t}) \otimes \mathcal{S}_G
\end{equation} where $\mathcal{S}$ is the quotient field of $H^*_{G\times \UU(1)^2}(pt)$.
A vector called the fundamental class $[M_{G,n,t}]$ is naturally defined  as an element in $H^*_{G\times \UU(1)^2}(M_{G,n,t})$. Then the vector $\ket{\text{pure}}$ above is \begin{equation}
\ket{\text{pure}}=\bigoplus_{n=0}^\infty [M_{G,n,t}] \in \bigoplus_{n=0}^\infty H^*_{G\times \UU(1)^2}(M_{G,n,t}).
\end{equation}

For general $G$, there is only the singular space $M_{G,n}$ and not the smooth version $M_{G,n,t}$. Still, using the equivariant intersection cohomology, one can write the partition function of the pure $\cN=2$ gauge theory with arbitrary gauge group $G$ in the form \eqref{purepartition}, see e.g.~\cite{BFN}.

\subsection{Physical interpretation of the reformulation}\label{physre}
The reformulation in the previous section can be naturally understood by considering a five-dimensional setup; it is important to distinguish it from another five-dimensional set-up we already used in Sec.~\ref{first5d}. 

Take the maximally supersymmetric $\SU(N)$ gauge theory with coupling constant $g$.
We put the system on $\bR^{1,3}$ times a segment in the $x^5$ direction, which is $[0,L_1] \cup [L_1,L_1+L_2] $.
We put boundary conditions at $x^5=0$, $L_1$ and $L_1+L_2$.  This necessarily breaks the supersymmetry to one half of the original, making it to a system with 4d $\cN=2$ supersymmetry. 
At $x^5=L_1$, we put $N\times N$ hypermultiplets, to which the $\SU(N)$ gauge group on the left and the $\SU(N)$ gauge group on the right couple by the left and the right multiplication. 
At $x^5=0$ and $x^5=L_1+L_2$, we put a boundary condition which just terminates the spacetime without introducing any hypermultiplet. See Fig.~\ref{higher} (5d).

In the scale larger than $L_{1,2}$, the theory effectively becomes the quiver gauge theory treated, because the segment $x^5\in [L_1+L_2,L_1]$ gives rise to an $\SU(N)$ gauge group with 4d gauge inverse square coupling $L_2/g^2_{5d}$, and the segment $[L_1,0]$ another $\SU(N)$ gauge group with 4d inverse square coupling $L_1/g^2_{5d}$.  Therefore we  have, in \eqref{quiverpartition}, \begin{equation}
\log q_1 / \log q_2 = L_1/ L_2.
\end{equation}

The final idea is to consider the $x^5$ direction as the time direction.
At each fixed value of $x^5$, one has a state in the Hilbert space of this quantum field theory,
which is $\cV_{G,\vec a}$  introduced in the previous section. Then every factor in the partition function of the quiver theory \eqref{quiverpartition} has a natural interpretation, see Fig.~\ref{higher} (eq):
\begin{itemize}
\item $\ket{\text{pure}}$ is the state created by the boundary condition at $x^5=0$.
\item  $q_1^\sN = e^{(\log q_1)\sN} = e^{- L_1 E}$ is the Euclidean propagation of the system by the length $L_1$.
\item $\Phi_{\vec a_2,m,\vec a_1}$ is the operation defined by the bifundamental hypermultiplet at $x^5=L_1$. 
\item $q_2^\sN=e^{(\log q_2)\sN} = e^{- L_2 E}$ is the Euclidean propagation of the system by the length $L_2$.
\item 
$\bra{\text{pure}}$ is the state representing the boundary condition at $x^5=L_1+L_2$.
\end{itemize}

\subsection{W-algebra action and the sixth direction}\label{WN}
For $G=\UU(N)$, it is a mathematical fact \cite{SchiffmannVasserot,MaulikOkounkov} that there is a natural action of the $W_N$ algebra on $\cV_{G,\vec a}$. 
The $W_N$ algebra is generated by two-dimensional holomorphic spin-$d$ currents $W_d(z)$, $d=2,3,\ldots, N$, and in particular contains the Virasoro subalgebra generated by $T(z)=W_2(z)$. 
The $L_0$ of the Virasoro subalgebra is identified with $\sN$ acting on $\cV_{G,\vec a}$. In particular, $L_{-m}$ maps $\cV_{G,\vec a,n}$  to $\cV_{G,\vec a,n+m}$. Figuratively speaking, $L_{-m}$ adds $m$ instantons into the system.
Furthermore, for generic value of $\vec a$, $\cV_{G,\vec a}$ is the Verma module of the $W_N$-algebra times a free boson. The central charge of the Virasoro subalgebra of this $W_N$ algebra is given by the formula \begin{equation}
c=(N-1) + N(N^2-1) \frac{(\epsilon_1+\epsilon_2)^2}{\epsilon_1\epsilon_2}.\label{cWN}
\end{equation}
Furthermore, it is believed that there is a natural decomposition\begin{equation}
\cV_{G,\vec a}=V_{\vec a'}\otimes H_m \label{tensor}
\end{equation}  into a $W_N$ Verma module $V_{\vec a'}$, and a free boson Fock space $H_m$.
Here, we define $m$ and $\vec a'$ via \begin{equation}
m=\sum a_i,\quad \vec a'=\vec a - m(\frac1N,\ldots,\frac1N).
\end{equation} Note that $\vec a'$ lives in an $N-1$ dimensional subspace. 
Then $V_{\vec a'}$ is the Verma module of the $W_N$ algebra constructed from $N-1$ free scalar fields with zero mode eigenvalue by $\vec a'$ and the background charge  \begin{equation}
\vec Q=(b+\frac 1b)(\frac N2,\frac N2-1,\ldots,1-\frac N2,-\frac N2),\quad b^2=\frac{\epsilon_1}{\epsilon_2},\label{bkg}
\end{equation}
and $H_m$ is the free boson Fock space with zero mode eigenvalue $m$.
The action of a free boson on $H_m$ was constructed in \cite{CarlssonOkounkov}. The decomposition above was also studied in  \cite{Fateev:2011hq,Belavin:2011sw}

When $\epsilon_1+\epsilon_2=0$, we have $b+1/b=0$ and the background charges \eqref{bkg} vanish. In this case the system becomes particularly simple, and  it was already studied in \cite{Losev:2003py,Aganagic:2003qj,Dijkgraaf:2007sw} 

The vector $\ket{\text{pure}}\in \cV_{G,\vec a}$, from this point of view, is a special vector called a Whittaker vector, which is a kind of a coherent state of the W-algebra \cite{Gaiotto:2009ma,Marshakov:2009gn,Taki:2009zd,Yanagida:2010kf}. Small number of hypermultiplets in the fundamental representation also is a boundary condition which also corresponds to a special state, studied in \cite{Kanno:2012xt}. 

The linear map $\Phi_{\vec a,m,\vec b}$ defined by a bifundamental hypermultiplet \eqref{bifop} should be a natural map between two representations of $W_N$ algebras. A natural candidate is an intertwiner of the $W_N$ algebra action, or equivalently, it is an insertion of a primary operator of $W_N$.
If that is the case, the partition function of a cyclic quiver with the gauge group $\SU(N)_1\times \SU(N)_2\times \SU(N)_3$, \begin{equation}
\tr q_1^\sN  \Phi_{\vec a,m_1,\vec b}  q_2^\sN \Phi_{\vec b,m_2,\vec c} q_3^\sN \Phi_{\vec c,m_3,\vec a}, 
\end{equation} for example, is the conformal block of the $W_N$ algebra on the torus $z\sim q_1q_2q_3z$ with three insertions at $z=1,$ $q_1$, and $q_1q_2$. This explains the observation first made in \cite{Alday:2009aq}.

Therefore the mathematically missing piece is to give the proof that $\Phi_{\vec a,m,\vec b}$ is the primary operator insertion. For $N=2$ when $W_N$ is the Virasoro algebra, this has been proven in \cite{Fateev:2009aw,Hadasz:2010xp}, but the general case is not yet settled. At least, there are many studies which show the agreement up to low orders in the $q$-expansion \cite{Mironov:2009dr,Mironov:2009by}. Also, the decomposition \eqref{tensor} predicts the existence of a rather nice basis in the Verma module of $W_N$ algebra times a free boson which was not know before, whose property was studied in \cite{Alba:2010qc}. The decomposition was also studied from the point of view of the $W_{1+\infty}$ algebra \cite{Kanno:2011qv,Kanno:2012hk} corresponding to the case $\epsilon_1+\epsilon_2=0$. Its generalization to the case $\epsilon_1+\epsilon_2\neq 0$ was done in \cite{Kanno:2013aha}.

When one considers a bifundamental charged under $\SU(N_1)\times \SU(N_2)$ with $N_1> N_2$, we have a linear operator  \begin{equation}
\Phi:\cV_{\SU(N_1)} \to \cV_{\SU(N_2)},
\end{equation} and we have an action of $W_{N_i}$ on $\cV_{\SU(N_i)}$.
The 6d construction using $\cN=(2,0)$ theory of type $\SU(N_1)$ \cite{Gaiotto:2009we} suggests that it can also be represented as a map \begin{equation}
\Phi:\cV_{\SU(N_1)} \to \cV',
\end{equation} where we still have an action of $W_{N_1}$ on $\cV'$. Then $\cV'$ is no longer a Verma module, even for generic values of parameters. $\cV'$ are believed to be the so-called  semi-degenerate representations of $W_{N_1}$ algebras determined by $N_2$, and there are a few checks of this idea \cite{Kanno:2009ga,Drukker:2010vg,Kanno:2010kj}.

\subsection{String theoretical interpretations}\label{top}

As seen in Sec.~\ref{physre}, the operator $\sN$ is the Hamiltonian generating the translation along $x^5$.  It is therefore most natural to make the identification $\log |z| = x^5$. Although the circle direction $x^6=\arg z$ was not directly present in the setup of Sec.~\ref{physre}, it also has a natural interpretation. Namely, the maximally supersymmetric 5d gauge theory with gauge group $\UU(N)$ on a space $X$ is in fact the six-dimensional $\cN=(2,0)$ theory of type $\UU(N)$ on a space $X\times S^1$, such that the Kaluza-Klein momentum along the $S^1$ direction is the instanton number of the 5d gauge theory. This again nicely fits with the fact that $L_n$ creates $n$ instantons, as the operator $L_n$  has $n$ Kaluza-Klein momenta along $S^1$. The quiver gauge theory treated at the end of Sec.~\ref{physre} can now be depicted as in Fig.~\ref{higher} (6d). There, the boundary conditions at both ends correspond to the state $\ket{\text{pure}}$ in $\cV$. The operator $\Phi_{\vec a,m,\vec b}$ is now an insertion of a primary field. 

If one prefers string theoretical language, it can be further rephrased as follows. We consider $N$ D4-branes on the space $X$, in a Type IIA set-up. This is equivalent to $N$ M5-branes on the space $X\times S^1$ in an M-theory set-up. The Kaluza-Klein momenta around $S^1$ are the D0-branes in the Type IIA description, which can be absorbed into the world-volume of the D4-branes as instantons. The insertion of a primary is an intersection with another M5-brane. This reduces in the type IIA limit an intersection with an NS5-brane, which gives the bifundamental hypermultiplet.

In the discussions so far, we introduced two vector spaces associated to the $n$-instanton moduli space $M_{G,n}$, and saw the appearance of \emph{three} distinct extra spacetime directions, $\xi^5$, $x^5$ and $x^6$.

\begin{itemize}
\item First, we introduced $\cH_n$ in Sec.~\ref{first5d}.  
We put $\cN=1$ supersymmetric 5d gauge theory with hypermultiplets on the $\Omega$ background $\bR^4\times S^1$ so that $\bR^4$ is rotated when we go around $S^1$. We then considered $S^1$ as the time direction. 
We called  this direction $\xi^5$.
The supersymmetric, non-perturbative part of the field theory Hilbert space reduces to the Hilbert space of the supersymmetric quantum mechanics on the moduli space of $n$-instantons plus the hypermultiplet zero modes.  We did not use the inner product in this Hilbert space. Mathematically, it is the space of holomorphic functions on the moduli space. 
\item Second, we introduced $\cV_n$ in Sec.~\ref{reformulation}.
We put the maximally-supersymmetric 5d gauge theory on $\bR^4\times$ a segment parameterized by $x^5$,  and considered  the segment as the time direction. The supersymmetric, non-perturbative part of the field theory Hilbert space reduces to the space $\cV_n$. It has an inner product, defined by means of the trace on $\cH_n$. Mathematically, $\cV_n$ is the equivariant cohomology of the moduli space. 
In this second setup, another circular direction $x^6$ automatically appears, so that it combines with $x^5$ to form a complex direction $\log z =x^5 + ix^6$.
\end{itemize}

It is important to keep in mind that in this second story with $x^5$ and $x^6$ we kept the radius $\beta$ of $\xi^5$ direction to be zero. If we keep it to a nonzero value instead, the inner product on $\cV_{G,\vec a}$ \eqref{4dinner} is instead modified to \begin{equation}
\vev{p|q}=\delta_{p,q} \frac{1}{\prod_{t} 1-e^{\ii \beta v(p)_t}}.\label{5dinner}
\end{equation} Let us distinguish the vector space with this modified inner product from the original one by calling it $\tilde\cV_{G,\vec a}$.  The $W_N$ action is no longer there. Instead, we have \cite{Awata:2009ur,Awata:2010yy,SchiffmannVasserot,MaulikOkounkov,Carlsson:2013jka} an action of $q$-deformed $W_N$ algebra on $\tilde \cV_{G,\vec a}$, which does not contain a Virasoro subalgebra. Therefore, we do not generate additional direction $x^6$ anymore. 
String theoretically, the set up with $\xi^5$ and $x^5$ corresponds to having $N$ D5-branes in Type IIB, and it is hard to add another physical direction to the system.

\begin{figure}[t]
\[
\begin{array}{ccccc}
(1) & (2)  & (3)  & (4) &(5) \\
\inc{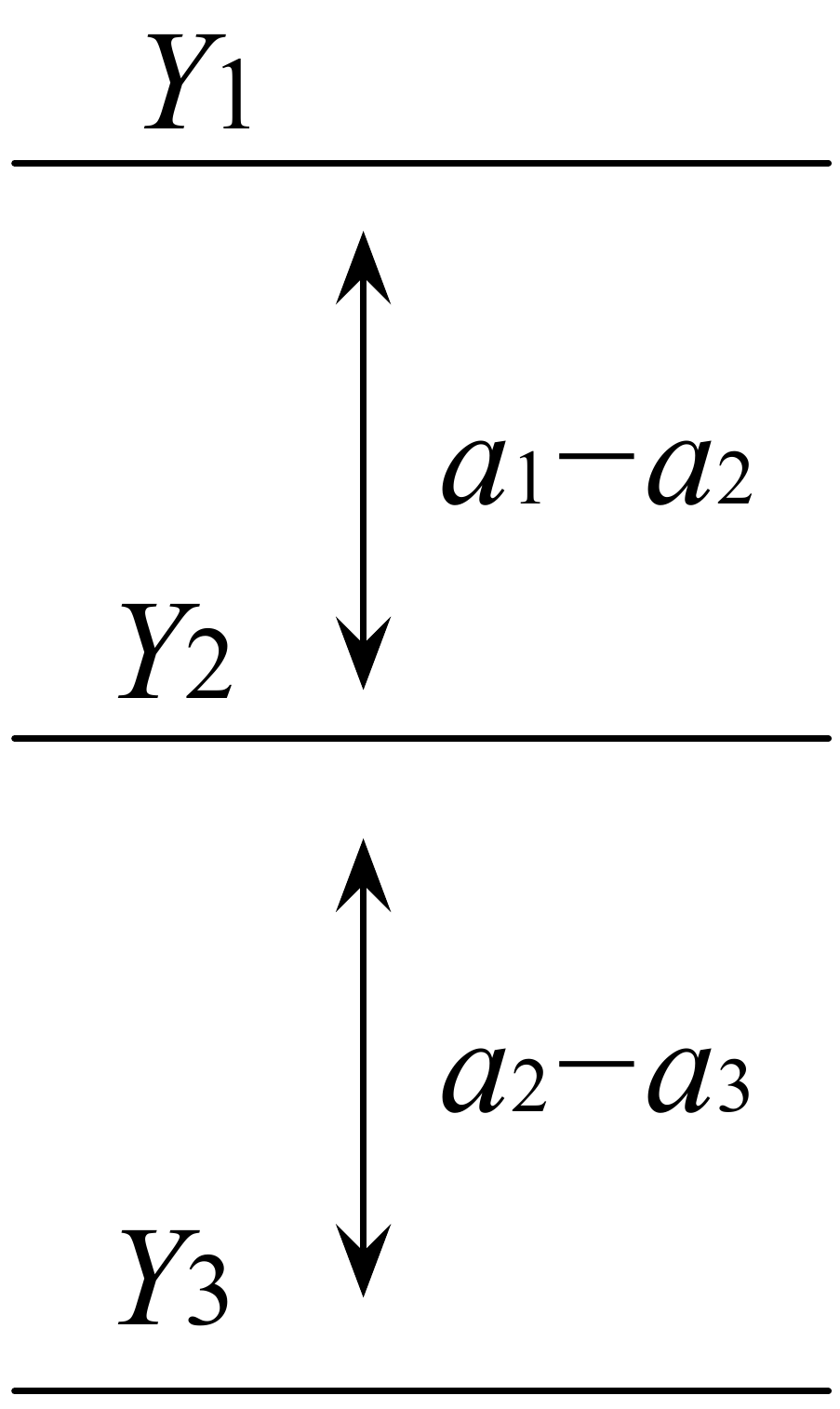} & \inc{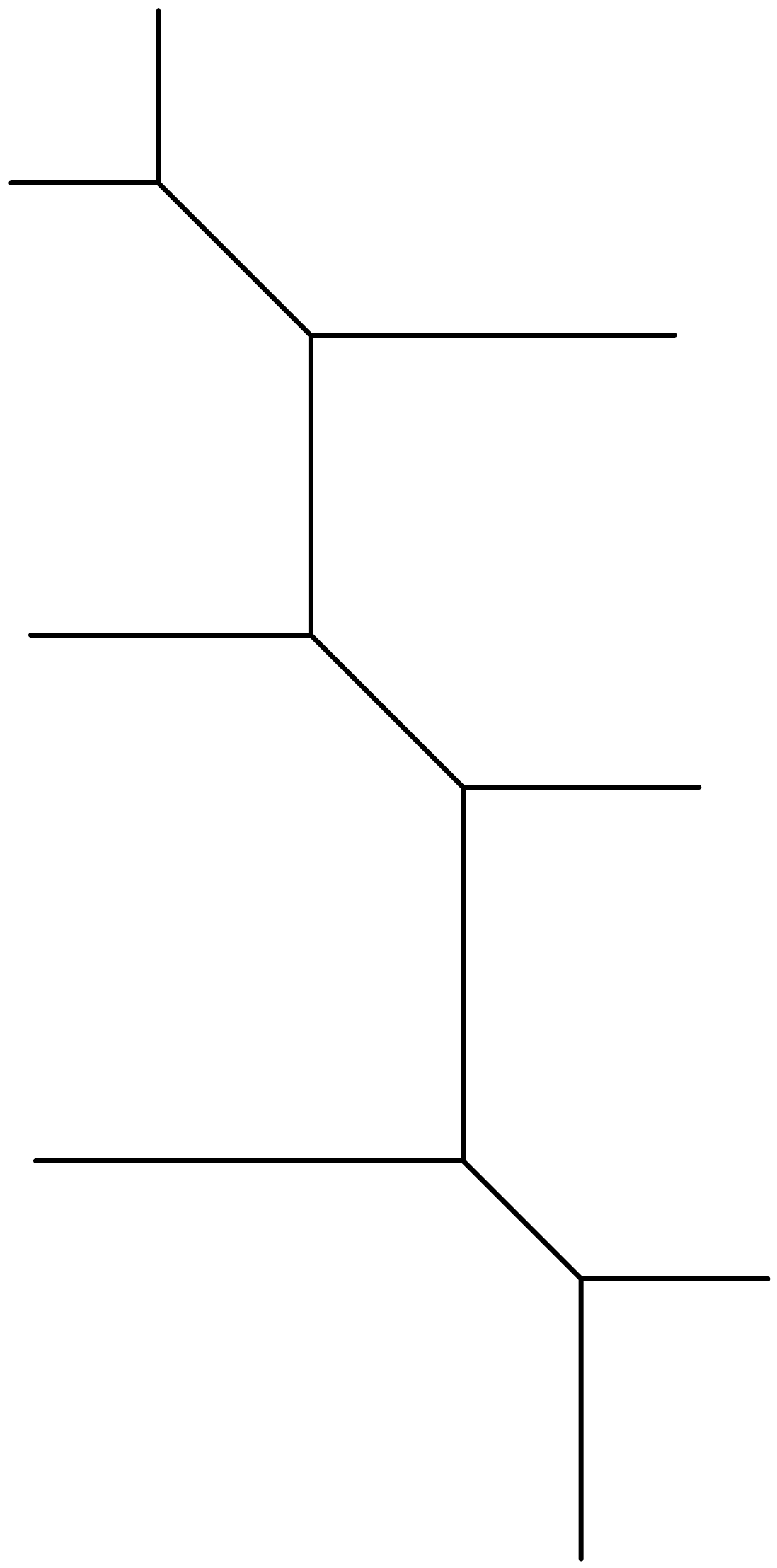} & \inc{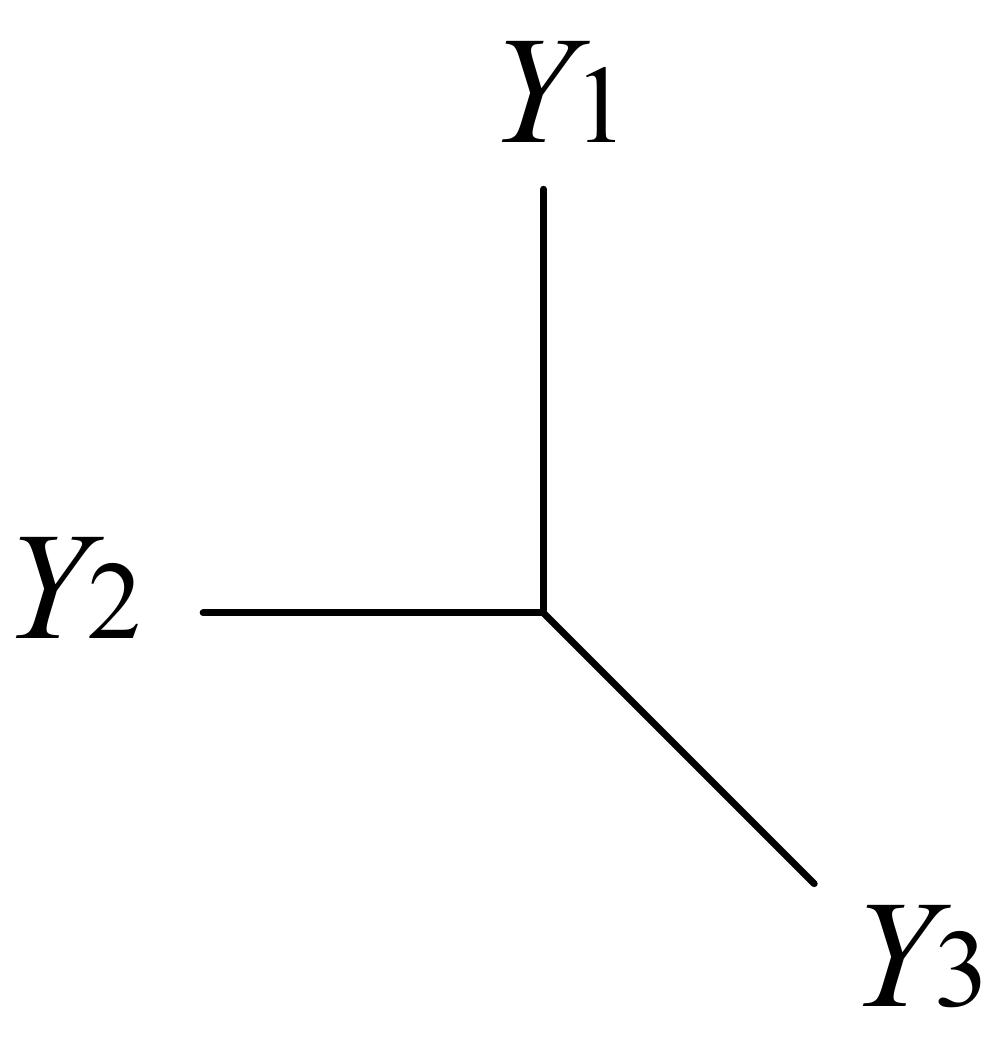} & \inc{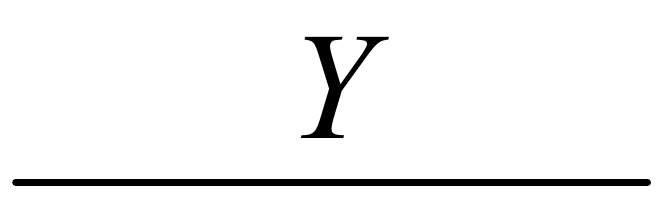} &\inc{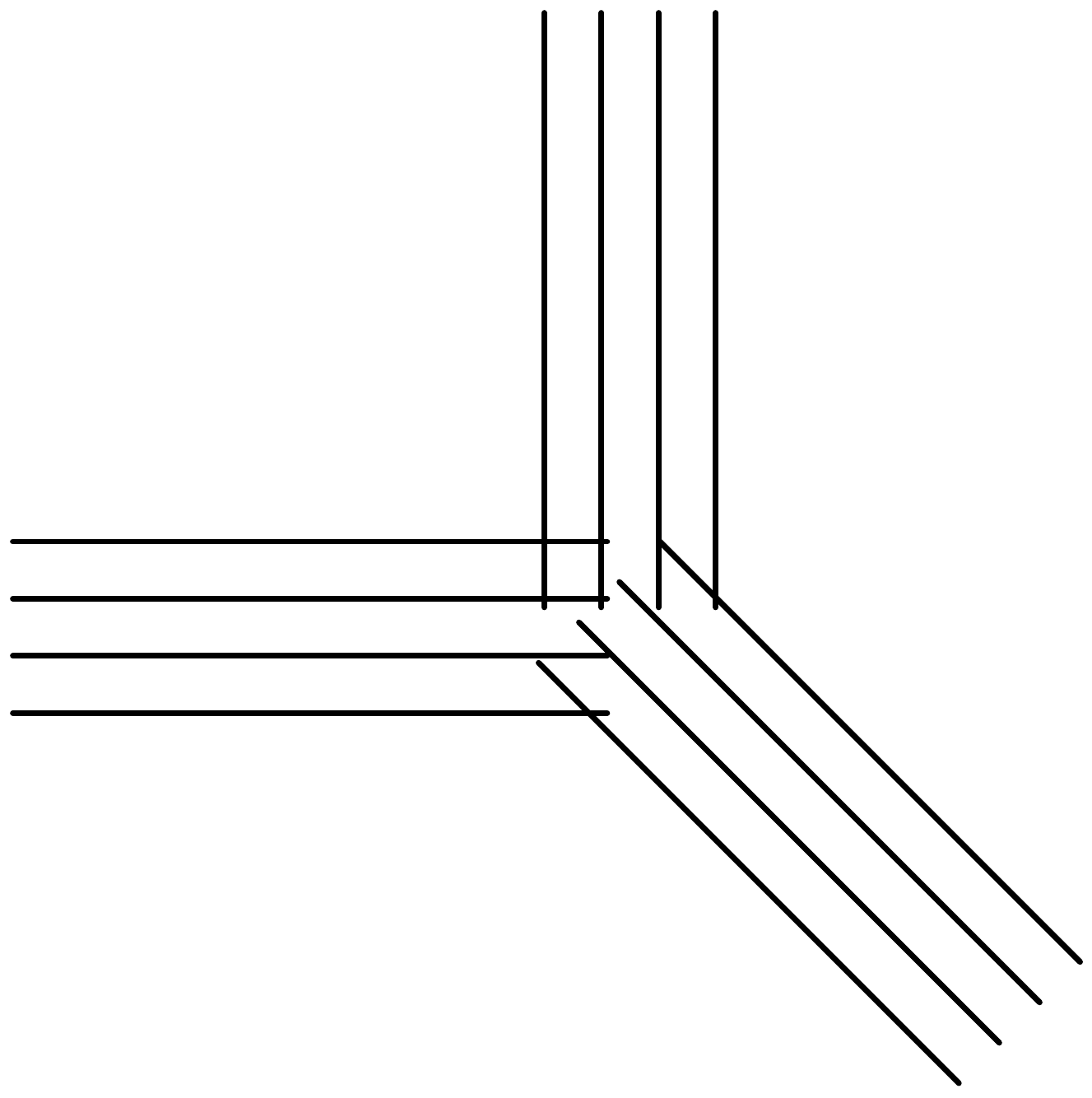}
\end{array}
\]
\caption{Type IIB setup, or equivalently the toric diagram.\label{tv}}
\end{figure}

\paragraph{Relation to the refined topological vertex}
Now, let us picturize this last Type IIB setup. We depict $N$ D5-branes as $N$  lines as in Fig~\ref{tv} (1). The horizontal direction is $x^5$, the vertical direction is $x^9$, say. We do not show the spacetime directions $\bR^4$ or the compactified direction $\xi^5$.  In the calculation of the instanton partition function, we assign a Young diagram to each D5-brane.

The boundary condition at fixed value of $x^5$, introducing a bifundamental hypermultiplet, is realized by an NS5-brane cutting across $N$ D5-branes, which can be depicted as in Fig~\ref{tv} (2).
When an NS5-brane crosses an D5-brane, they merge to form a (1,1) 5-brane, which needs to be tilted to preserve supersymmetry; the figure shows this detail. 

Therefore, the whole brane set-up describing a five-dimensional quiver gauge theory on a circle can be built from a vertex joining three 5-branes Fig~\ref{tv} (3), and a line representing a 5-brane Fig.~\ref{tv} (4). Any 5-brane is obtained by an application of the $\SL(2,\bZ)$ duality to the 5-brane, so one can associate a Young diagram to any line. The basic quantity is then a function $Z_\text{vertex}(\epsilon_1,\epsilon_2;Y_1,Y_2,Y_3)$ which is called the refined topological vertex.
The partition function of the system is obtained by multiplying the refined topological vertex for all the junctions of three 5-branes, multiplying a propagator factor $\Delta(Y)$ for each of the internal horizontal line, and summing over all the Young diagrams. 

The phrase `refined topological' is used due to the following situation where it was originally discovered. A  review of the detail can be found in \cite{A} in this volume, so we will be brief here.
We  apply  a further chain of dualities to the setup we have arrived, so that the diagrams in Fig.~\ref{tv} are now considered as specifying the toric diagram of a non-compact toric Calabi-Yau space on which M-theory is put. The direction $\xi^5$ is now the M-theory circle. 
Nekrasov's partition function of this setup when $\epsilon_1=-\epsilon_2=g_s$ is given by the partition function of the topological string on the same Calabi-Yau with the topological string coupling constant at $g_s$. This gives the unrefined version of the topological vertex. 
The generalized case $\epsilon_1\neq -\epsilon_2$ should correspond to a refined version of the topological string on the Calabi-Yau, and the function $Z_\text{vertex}$ for general $\epsilon_1$, $\epsilon_2$ is called the refined topological vertex. The unrefined version was determined in \cite{Iqbal:2002we,Aganagic:2003db} and the refined version was determined in \cite{Awata:2005fa,Iqbal:2007ii,Iqbal:2012mt}. 

In this discussion, we implicitly used the fact that the logarithm of the partition function on the $\Omega$ background \eqref{omegaback} is  equal to the prepotential in the presence of the graviphoton background, which is further equal to the free energy of the topological string.  In the unrefined case this identification goes back to \cite{Antoniadis:1993ze,Bershadsky:1993cx}. The refined case is being clarified, see e.g.~\cite{Antoniadis:2010iq,Nakayama:2010ff,Nakayama:2011be}.

As an aside, we can also perform a T-duality along the $\xi_5$ direction in the type IIB configuration above. This gives rise to a type IIA configuration in the fluxtrap solution, which lifts to a configuration of M5-branes with four-form background \cite{Hellerman:2011mv,Reffert:2011dp,Hellerman:2012zf}.
For a certain class of gauge theories, we can also go to a duality frame where we have  D3-branes in an orbifold singularity with a particular RR-background.  It has been directly checked that the partition function in this setup reproduces Nekrasov's partition function in the unrefined case $\epsilon_1+\epsilon_2=0$ \cite{Billo:2006jm,Ito:2008qe,Ito:2010vx}.

Let us come back to the discussion of the  refined topological vertex itself. The summation over the Young diagrams in the internal lines of Fig.~\ref{tv} (2) can be carried out explicitly using the properties of Macdonald polynomials, and correctly reproduces the numerator of the partition function  \eqref{5dpartition} coming from a bifundamental, given by the weights in \eqref{bif}. The denominator basically comes from the propagator factors associated to $N$ horizontal lines \cite{Iqbal:2004ne,Taki:2007dh}.

Here, it is natural to consider an infinite dimensional vector space \begin{equation}
\tilde\cV_1 = \bigoplus_{Y} \bC \ket{Y}
\end{equation} whose basis is labeled by a Young diagram, such that the inner product is given by the propagator factor $\Delta(Y)$ of the topological vertex. 
Now, the space $\tilde\cV_1$ is known to have a natural action of an algebra called the Ding-Iohara algebra $\cDI$  \cite{Awata:2011dc}.  It might be helpful to know that this algebra is also called the elliptic Hall algebra, or the quantum toroidal $\GL(1)$ algebra; see e.g.~\cite{GKV} for the quantum toroidal algebras.
Then the refined topological vertex $Z_\text{vertex}$ is an intertwiner of this algebra: \begin{equation}
Z_\text{vertex}: \tilde\cV_1 \otimes \tilde\cV_1 \to \tilde\cV_1.\label{1}
\end{equation} The $q$-deformed $W_N$-algebra action on $\tilde\cV_{\UU(N)}$, from this point of view, should be understood from its relation to the action of the Ding-Iohara algebra $\cDI$ on 
\begin{equation}
\tilde\cV_1{}^{\otimes N} \simeq \tilde\cV_{\UU(N)}.\label{map}
\end{equation}
The $W_N$ action on $\cV_{\UU(N)}$ should follow when one takes the four-dimensional limit when the radius $\beta$ of the $\xi^5$ direction goes to zero. 

This formulation has an advantage that the instanton partition function on $S^1$ of a 5d non-Lagrangian theory, such as the $T_N$ theory corresponding to Fig.~\ref{tv} (5), can be computed, by just multiplying the vertex factors and summing over Young diagrams. 
Indeed this computation was performed in \cite{Bao:2013pwa,Hayashi:2013qwa}, where the $E_6$ symmetry of the partition function of $T_3$ was demonstrated. 

It should almost be automatic that the resulting partition function $Z_{T_N}$ of $T_N$ is an intertwiner of $q$-deformed $W_N$ algebra, because the linear map 
\begin{equation}
Z_{T_N}: \tilde\cV_{\UU(N)} \otimes \tilde\cV_{\UU(N)} \to \tilde\cV_{\UU(N)}
\end{equation} is obtained by composing $N(N-1)/2$ copies of $Z_\text{vertex}$ according to Fig.~\ref{tv} (5). 
One can at least hope that the intertwining property of $Z_\text{vertex}$, together with the naturality of the map \eqref{map}, should translate to the intertwining property of $Z_{T_N}$.

\section{Other gauge groups}\label{othergauge}
In this section, we indicate how the instanton calculations can be extended to gauge groups other than (special) unitary groups.
We do not discuss the details, and only point to the most relevant results in the literature. 
\subsection{Classical gauge groups}\label{otherclassical}
 Let us consider classical gauge groups $G=\SO(2n)$, $\SO(2n+1)$ and $\USp(2n)$. Physically, nothing changes from what is stated in Sec.~\ref{first5d}; we need to perform localization on the $n$-instanton moduli space $M_{G,n}$ of gauge group $G$. A technical problem is  that there is no known way to resolve and/or deform the singularity of $M_{G,n}$ to make it smooth, when $G$ is not unitary. 

To proceed, we first re-think the way we performed the calculation when $G=\UU(N)$.  For classical $G$, the instanton moduli space has the ADHM description, just as in the unitary case recalled in \eqref{ADHM}: \begin{equation}
M_{G,n}= \{  \mu_{\bC}(x)=0 \mid x\in X_{G,n}  \} / K(G,n)
\end{equation} Here, $K(G,n)$ is a complexified compact Lie group, and $X_{G,n}$ is a vector space, given as in \eqref{Xn} by a tensor product and a direct sum starting from vector spaces $V$ and $W$  which are the fundamental representations of $K(G,n)$ and $G$ respectively.
One can formally rewrite the integral which corresponds to the localization on $M_{G,n}$ as an integral over \begin{equation}
X_{G,n} \oplus \fk_\bR(G,n),
\end{equation} where $\fk(G,n)$ is the Lie algebra of $K(G,n)$. The integral along $X_{G,n}$ can be easily performed, and the integration on $\fk_\bR(G,n)$ can be reduced to an integration on the Cartan subalgebra $\fh_\bR(G,n)$ of $\fk_\bR(G,n)$, resulting in a formal expression \begin{equation}
Z_{\text{inst},n}(\epsilon_{1,2};a_{1,\ldots,r};m_{1,\ldots, f}) = \int_{\phi\in \fh_\bR(G,n)} 
f(\epsilon_{1,2};a_{1,\ldots,r};m_{1,\ldots, f};\phi)
\label{4dpartition-classical}
\end{equation} where $f$ is a rational function. 

The fact that $M_{G,n}$ is singular is reflected in the fact that the poles of the rational function are on the integration locus $\fh_\bR(G,n)$. When $G$ is unitary, the deformation of the instanton moduli space $M_{G,n}$ to make it smooth corresponds to a systematic deformation of the half-dimensional integration contour $\fh_\bR(G,n)\subset \fh_\bC(G,n)$. Furthermore, the poles are in one-to-one correspondence with the fixed points on the smoothed instanton moduli space.
A pole is given by a specific value $\phi\in \fh_\bC(G,n)$ which is a certain linear combinations of $\epsilon_{1,2}$, $a_{1,\ldots,r}$, and $m_{1,\ldots, f}$. In other words, the position of a pole is given by specifying the action of $\UU(1)^{2+r} \subset \UU(1)^2\times G$  on the vector space $V$, which is naturally a representation of $K(G,n)$. 
Finally, the residues give the summand in the localization formula \eqref{4dpartition}. 

Although the deformation of the moduli space is not possible when $G$ is not unitary, the systematic deformation of the integration contour $\fh_\bR(G,n)\subset \fh_\bC(G,n)$ is still possible. The poles are still specified by the actions \begin{equation}
\phi_p : \UU(1)^{2+r}  \curvearrowright  V.
\end{equation}
Then the instanton partition function \eqref{4dpartition-classical} can be written down explicitly as \begin{equation}
Z_{\text{inst},n}(\epsilon_{1,2};a_{1,\ldots,r};m_{1,\ldots, f}) = \sum_p \mathrm{Res}_{\phi=\phi_p} f(\epsilon_{1,2};a_{1,\ldots,r};m_{1,\ldots, f};\phi).
\end{equation}
This calculation was pioneered in \cite{Marino:2004cn,Nekrasov:2004vw}, and further elaborated in \cite{Hollands:2010xa}. 

\subsection{Effect of finite renormalization}
Let us in particular consider $\cN=2$ $\SU(2)$ gauge theory with four fundamental hypermultiplets, with  all the masses set to zero for simplicity. Its instanton partition function can be calculated either as the $N=2$ case of $\SU(N)$ theory, or as the $N=1$  case of $\USp(2N)$ theory, using the ADHM construction either of the $\SU$ instantons or of the $\USp$ instantons.
What was found in \cite{Hollands:2010xa} is that the $n$-instanton contribution calculated in this manner, are all different: \begin{equation}
Z_{\text{inst},n}^{\SU(2)}(\epsilon_{1,2};a) \neq 
Z_{\text{inst},n}^{\USp(2)}(\epsilon_{1,2};a).
\end{equation} They also found that the total instanton partition functions \begin{equation}
Z^G(q;\epsilon_{1,2};a) = q^{a^2/\epsilon_1\epsilon_2} \sum_{n\ge 0} q^n Z_{\text{inst},n}^{G}(\epsilon_{1,2};a;m)
\end{equation} becomes the same, \begin{equation}
Z^{\SU(2)}(q_{\SU(2)};\epsilon_{1,2};a;m)=
Z^{\USp(2)}(q_{\USp(2)};\epsilon_{1,2};a;m)\label{regindependence}
\end{equation} once we set \begin{equation}
q_{\SU(2)}=q_{\USp(2)} ( 1 + \frac{q_{\USp(2)}}4)^{-2}.\label{finiterenormalization}
\end{equation}

The physical coupling $q_\text{IR}=e^{2\pi \tau_\text{IR}}$ in the infrared is then given in terms of the prepotential: \begin{equation}
2\pi \tau_\text{IR}  \lim_{\epsilon_{1,2}\to 0} \epsilon_1\epsilon_2\log Z^G(q;\epsilon_{1,2}). 
\end{equation} This is given by \begin{equation}
q_{\SU(2)}=\frac{\theta_2(q_\text{IR})^4}{\theta_3(q_\text{IR})^4}\label{IR}
\end{equation} This finite discrepancy between the UV coupling $q_{\SU(2)}$ and the IR coupling $q_\text{IR}$ was first clearly recognized in \cite{Dorey:1996ez}, and the all order form was conjectured by \cite{Huang:2010kf}. We see that the UV coupling $q_{\USp(2)}$ is different from both. 

These subtle difference among $q_{\SU(2)}$, $q_{\USp(2)}$ and $q_\text{IR}$ reflects a standard property of any well-defined quantum field theory. The factor weighting the instanton number, $q_G$, is an ultra-violet dimensionless quantity, and is renormalized, the amount of which depends on the regularization chosen. The choice of the ADHM construction of the $\SU(2)=\USp(2)$ instanton moduli space and the subsequent deformation of the contours are part of the regularization. The final physical answer should be independent \eqref{regindependence}, once the finite renormalization is correctly performed, as in \eqref{finiterenormalization}. 

In this particular case, there is a natural geometric understanding of the relations \eqref{finiterenormalization} and \eqref{IR} \cite{Hollands:2010xa}. 
The $\SU(2)$ theory with four flavors can be realized by putting 2 M5-branes on a sphere $C$ with four punctures $a_1,a_2,b_1,b_2$, whose cross ratio is the UV coupling $q_{\SU(2)}$.  
The Seiberg-Witten curve of the system is  the elliptic curve $E$ which is a double-cover of $C$ with four branch points at $a_1,a_2,b_1,b_2$.  
The IR gauge coupling is the complex structure of $E$, and this gives the relation \eqref{IR}.

The same system can be also realized by putting 4 M5-branes on top of the M-theory orientifold 5-plane on a sphere $C'$ with four punctures, $x,y,a,b$, whose cross ratio is the coupling $q_{\USp(2)}$. Here we also have the orientifold action around the puncture $x$, $y$. There is a natural 2-to-1 map $C\to C'$ with branch points at $x$ and $y$, so that $a_{1,2}$ and $b_{1,2}$ on $C$ are inverse images of $a$ and $b$ on $C'$, respectively. This gives the relation \eqref{finiterenormalization}.

\subsection{Exceptional gauge groups}\label{exceptional}
For exceptional gauge groups $G$ , not much was known about the instanton moduli space $M_{G,n}$, except at instanton number $n=1$, because we do not have ADHM constructions. 
To perform the instanton calculation in full generality in the presence of matter hypermultiplets, we need to know the properties of various bundles on $M_{G,n}$.  For  the pure gauge theory, the knowledge of the ring of the holomorphic functions on $M_{G,n}$ would suffice. Any instanton moduli space decomposes as $M_{G,n}=\bC^2\times M_{G,n}^\text{centered}$, where $\bC^2$ parameterize the center of the instanton, and $M_{G,n}^\text{centered}$ is called the centered instanton moduli space. Therefore the question is to understand the centered instanton moduli space better. 

The centered one-instanton moduli space of any gauge group $G$ is the minimal nilpotent orbit of $\fg_{\bC}$, i.e. the orbit under $G_\bC$ of a highest weight vector. The ring of the holomorphic functions on the minimal nilpotent orbit is known \cite{VinbergPopov,Garfinkle,Benvenuti:2010pq}, and thus the instanton partition function of pure exceptional gauge theory can be computed up to instanton number 1 \cite{Keller:2011ek}.

There are 4d $\cN=2$ quantum field theories ``of class S'' whose Higgs branch is $M_{E_r,n}^\text{centered}$ \cite{Benini:2009gi}. There is now a conjectured formula which computes the ring of holomorphic functions on the Higgs branch of a large subclass of class S theories \cite{Gadde:2011uv}. A review can be found in \cite{RR} in this volume.  This method can be  used to study $M_{E_r,2}$ explicitly, from which the instanton partition function of $E$-type gauge theories can be found \cite{Gaiotto:2012uq,Hanany:2012dm,Keller:2012da}. 

Moreover, the Higgs branch of any theories of class S is obtained \cite{Moore:2011ee} by the hyperk\"ahler modification \cite{Bielawski} of the Higgs branch of the so-called $T_G$ theory. The Higgs branch of the $T_G$ theory is announced to be rigorously constructed \cite{GinzburgKazhdan}. Therefore, we now have a finite-dimensional construction of $M_{E_r,n}$. This should allow us to perform any computation on the instanton moduli space, at least in principle.

\subsection{Relation to W-algebras}
We can form an infinite-dimensional vector space $\cV_G$ as in Sec.~\ref{reformulation}. 
When $G=\SU(N)$, there was an action of the $W_N$ algebra on $\cV_G$. There is a general construction of W-algebras starting from arbitrary  affine Lie algebras $\hat G$ and twisted affine Lie algebras $\hat G^{(s)}$ where $s=2,3$ specifies the order of the twist; in this general notation, the $W_N$ algebra is $W(\widehat{\SU(N)})$ algebra. For a comprehensive account of W-algebras, see the review  \cite{Bouwknegt:1992wg} and the reprint volume \cite{Bouwknegt:1995ag}.

When $G$ is simply-laced, i.e.~$G=\SU(N)$, $\SO(2N)$ or $E_N$, $\cV_G$ has an action of the $W(\hat G)$ algebra; this can be motivated from the discussion as in Sec.~\ref{WN}. We start from the 6d $\cN=(2,0)$ theory of type $G$, and put it on $\bR^4\times C_2$ where $C_2$ is a Riemann surface, so that we have $\cN=2$ supersymmetry in four dimensions. Then, we should have some kind of two-dimensional system on $C_2$. The central charge of this two-dimensional system can be computed \cite{Bonelli:2009zp,Alday:2009qq} starting from the anomaly polynomial of the 6d theory, which results in \begin{equation}
c=\rank G+ h^\vee(G)\dim G \frac{(\epsilon_1+\epsilon_2)^2}{\epsilon_1\epsilon_2}. 
\end{equation} This is the standard formula of the central charge of the $W(\hat G)$ algebra, when $G$ is simply-laced. 

\begin{table}[h]
\[
\begin{array}{c|cccc}
\Gamma & A_{2n-1} &  D_{n+1} & D_4 &  E_6   \\
\hline
s & 2 & 2 &  3 & 2  \\
\hline
G & B_n & C_n & G_2 & F_4 
\end{array}
\]
\caption{The type of the 6d theory, the choice of the outer-automorphism twists, and the 5d gauge group\label{twist-table}}
\end{table}

When $G$ is not simply-laced, we can use the physical 5d construction in Sec.~\ref{physre}, but there is no 6d $\cN=(2,0)$ theory of the corresponding type. Rather, one needs to pick a simply-laced $J$ and a twist $\sigma$ of order $s=2,3$, such that the invariant part of $J$ under $\sigma$ is Langlands dual to $G$, see the Table~\ref{twist-table}. Then, the 5d maximally supersymmetric theory with gauge group $G$ lifts to a 6d theory of type $J$, with the twist by $\sigma$ around $x^6$. This strongly suggests that the W-algebra which acts on $\cV_G$ is $W(\hat G^{(s)})$.  This statement was checked to the one-instanton level in \cite{Keller:2011ek} by considering pure $G$ gauge theory. A full mathematical proof for simply-laced $G$ is available in \cite{Braverman:2014xca}.

\section{Other spaces}\label{otherspace}
\subsection{With a surface operator}\label{surface}
\paragraph{Generalities}
Let us consider a gauge theory with a simple gauge group $G$, with a surface operator supported on $\bC\subset \bC^2$. A detailed review can be found in \cite{G}, so we will be brief here.   A surface operator is defined in the path integral formalism as in the case of 't Hooft loop operators, by declaring that fields have  prescribed singularities there. In our case, we demand that the gauge field has the divergence \begin{equation}
A_\theta d \theta \to \mu d\theta \label{sing}
\end{equation} where $\theta$ is the angular coordinate in the plane transverse to the surface operator, $\mu$ is an element in $\fg$; the behavior of other fields in the theory is set so that the surface operator preserves a certain amount of supersymmetry.  

On the surface operator, the gauge group is broken to a subgroup $L$ of $G$ commuting with $\mu$. 
Let us say there is a subgroup $\UU(1)^k\subset L$. Then, the restriction of the gauge field on the surface operator can have nontrivial monopole numbers $n_1,\ldots,n_k$. Together with the instanton number $n_0$ in the bulk, they comprise a set of numbers classifying the topological class of the gauge field. Thus we are led to consider the moduli space $M_{G,L,\mu,n_0,n_1,\ldots,n_k}$.
It is convenient to redefine $n_0,\ldots,n_k$ by an integral linear matrix  so that these instanton moduli spaces are nonempty if and only if $n_0,\ldots, n_k\ge 0$.
The instanton partition function is schematically given by \begin{equation}
Z_\text{inst}(\epsilon_{1,2};a_i;q_{0,1,\ldots,k})=
\sum_{n_0,n_1,\ldots,n_k\ge 0} q_0^{n_0}\cdots q_k^{n_k} Z_{\text{inst},n_0,\ldots,n_k}(\epsilon_{1,2};a_i)
\end{equation} where $Z_{\text{inst},n_0,\ldots,n_k}(\epsilon_{1,2};a_i)$ is given by a geometric quantity associated to $M_{G,L,\mu,n_0,n_1,\ldots,n_k}$.

This space is not well understood unless $G$ is unitary. Suppose $G$ is $\SU(N)$ . Then the singularity is specified by \begin{equation}
\mu=\diag(\underbrace{\mu_1,\ldots,\mu_1}_{m_1},\underbrace{\mu_2,\ldots,\mu_2}_{m_2},
\ldots,\underbrace{\mu_{k+1},\ldots,\mu_{k+1}}_{m_{k+1}}).\label{singg}
\end{equation}  Then the group $L$ is \begin{equation}
L=\mathrm{S}[\prod_{i=1}^{k+1} \UU(m_i)]\label{SUL}
\end{equation} which has a $\UU(1)^k$ subgroup. 

Here, we can use a mathematical result \cite{MehtaSeshadri,Biswas} which says that the moduli space $M_{G,L,\mu,n_0,n_1,\ldots,n_k}$ in this case is equivalent as a complex space to the moduli space of instantons on an orbifold $\bC/\bZ_{k+1} \times \bC$. As we will review in the next section, the instanton moduli space on an arbitrary Abelian orbifold of $\bC^2$  can be easily obtained from the standard ADHM construction, resulting in the quiver description of the instanton moduli space with a surface operator \cite{FR,FFNR}. The structure of the fixed points can also be obtained starting from that of the fixed points on $\bC^2$.  Then the instanton partition function can be explicitly computed \cite{Alday:2010vg,Kozcaz:2010yp}, although the details tend to be rather complicated when $[m_1,\ldots,m_{k+1}]$ is generic \cite{Wyllard:2010rp,Wyllard:2010vi,Kanno:2011fw}.

\paragraph{Corresponding W-algebra}
An infinite dimensional vector space $\cV_{G,L,\vec a}$ can be introduced as in Sec.~\ref{reformulation}: \begin{equation}
\cV_{G,L,\vec a}=\bigoplus_{n_0,\ldots, n_k\ge 0} \cV_{G,L;\vec a;n_0,\ldots, n_k}\label{VGL}
\end{equation} where $\cV_{G,L;\vec a;n_0,\ldots, n_k}$ is the equivariant cohomology of $M_{G,L,\mu,n_0,n_1,\ldots,n_k}$ with the equivariant parameter of $\SU(N)$ given by $\vec a$. As $\cV$ does not depend on the continuous deformation of $\mu$ with fixed $L$, we dropped $\mu$ from the subscript of $\cV$.

The W-algebra which is believed to be acting on $\cV_{G,L}$ is obtained as follows, when $G=\SU(N)$ and $L$ is given as in \eqref{SUL}.
Introduce an $N$-dimensional representation of $\SU(2)$ \begin{equation}
\rho_{[m_1,\ldots,m_{k+1}]} : \SU(2)\to \SU(N)
\end{equation} such that the fundamental representation of $\SU(N)$ decomposes as the direct sum of $\SU(2)$ irreducible representations with dimensions $m_1$, \ldots, $m_{k+1}$. 
Let us define a nilpotent element via \begin{equation}
\nu_{[m_1,\ldots,m_{k+1}]} =\rho_{[m_1,\ldots,m_{k+1}]} (\sigma^+).
\end{equation} 
Then we perform the quantum Drinfeld-Sokolov reduction of $\hat G$ algebra via this nilpotent element, which gives the algebra $W(\hat G,\nu_{[m_1,\ldots,m_{k+1}]})$ which is what we wanted to have.
In particular, when $[m_1,\ldots,m_{k+1}]=[1,\ldots,1]$,  the nilpotent element is $\nu=0$, and the resulting W-algebra is $\hat G$. When $[m_1,\ldots,m_{k+1}]=[N]$, there is no singularity, and the W-algebra is the standard $W(\hat G)$ algebra. 
The general W-algebras $W(\hat G,\nu)$ were introduced in \cite{deBoer:1993iz}.

Let $F$ be the commutant of $\rho_{[m_1,\ldots,m_{k+1}]}(\SU(2))$ in $\SU(N)$. Explicitly, it is \begin{equation}
F=\mathrm{S}[\prod_{s=1}^t \UU(\ell_t)] \label{fsurface}
\end{equation} where $\ell_{1,\ldots,t}$  is defined by writing \begin{equation}
[m_1,\ldots,m_k] = [n_1{}^{\ell_1},\ldots,n_t{}^{\ell_t}].
\end{equation} Note that the rank of $F$ is $k$.
The W-algebra $W(\hat G,\nu_{[m_1,\ldots,m_{k+1}]})$  contains an affine subalgebra $\hat F$. Therefore, the dimension of the Cartan subalgebra of $W(\hat G,\nu_{[m_1,\ldots,m_{k+1}]})$ is $\rank F+1=k+1$, and any representation of the W-algebra is graded by integers $n_0,\ldots,n_k$.  This matches with the fact that $\cV_{G,L}$ is also graded by the same set of integers \eqref{VGL}.

\paragraph{Higher-dimenisonal interpretation} 
From the 6d perspective advocated in Sec.~\ref{physre}, one considers a codimension-2 operator of the 6d $\cN=(2,0)$ theory of type $\SU(N)$, extending along $x^5$ and $x^6$. Such a codimension-2 operator is labeled by a set of integers $[m_1,\ldots,m_k]$, and is known to create a singularity of the form \eqref{sing}, \eqref{singg} in the four-dimensional part \cite{Gaiotto:2009we,Chacaltana:2010ks}. Furthermore, the operator is known to have a flavor symmetry $F$ as in \eqref{fsurface}. Therefore, it is as expected that the W-algebra $W(\hat G,\nu_{[m_1,\ldots,m_{k+1}]})$ has the $\hat F$ affine subalgebra. Its level can be computed by starting from the anomaly polynomial of the codimension-2 operator; a few checks of this line of ideas  were performed in \cite{Tachikawa:2011dz,Kanno:2011fw,Wyllard:2011mn}.

The partition function with surface operator of type $[N-1,1]$ can also be represented as an insertion of a degenerate primary field $\Phi$ in the standard $W_N$ algebra \cite{Drukker:2009id,Alday:2009fs,Kozcaz:2010af}.
When $N=2$, we therefore have two interpretations: one is that the surface operator  changes the Virasoro algebra to $\widehat{\SU(2)}$, the other is that the surface operator is a degenerate primary field of the Virasoro algebra. These can be related by the Ribault-Teschner relation \cite{Ribault:2005wp,Hikida:2007tq}, but the algebraic interpretation is not clear.

For general simply-laced $G$ and $L$, the  W-algebra which acts on $\cV_{G,L}$ is thought to be $W(\hat G,\nu)$, where $\nu$ is a generic nilpotent element in $L$. But there is not many explicit checks of this general statement, except when $L$ is the Cartan subgroup.

\paragraph{Braverman-Etingof} When $L$ is the Cartan subgroup, $\nu=0$, and the W-algebra is just the $\hat G$ affine algebra. 
Its action on $\cV_{G,L}$ was constructed in \cite{Braverman:2004vv}. The instanton partition function $Z$ of the pure $G$ gauge theory with this surface operator was then analyzed in \cite{Braverman:2004cr}. The limit \begin{equation}
F=\lim_{\epsilon_{1,2}\to 0}\epsilon_1\epsilon_2\log Z
\end{equation}
was shown to be independent of the existence of the surface operator; the surface operator contributes only a term of order $1/\epsilon_1$ to $\log Z$ at most.  The  structure of the $\hat G$ affine Lie algebra was then used to show that $F$ is the prepotential of the Toda system of type $G$, thus proving that the instanton counting gives the same prepotential as determined by the Seiberg-Witten curve.

Before proceeding, let us consider the contribution from the bifundamental hypermultiplet.
Again as in Sec.~\ref{reformulation}, it determines a nice linear map \begin{equation}
\Phi_{\vec a,m,\vec b}:\cV_{G,L,\vec a} \to \cV_{G,L,\vec b}
\end{equation} where $m$ is the mass of the hypermultiplet. This $\Phi_{\vec a,m,\vec b}$ is expected to be a primary operator insertion of this W-algebra. This is again proven when $\nu=0$ and the W-algebra is just the $\hat G$ affine algebra \cite{Negut2}. 

The author does not know how to incorporate hypermultiplet matter fields in this approach.

\subsection{On orbifolds}
Let us now consider the moduli space of instantons on an orbifold of $\bC^2$ by the $\bZ_p$ action \begin{equation}
g: (z,w)\to (e^{2\pi \ii  s_1/p}z,e^{2\pi \ii s_2/p}w).\label{gg}
\end{equation} 
This was analyzed by various groups, e.g.~\cite{Bonelli:2012ny,Bruzzo:2013daa}.
We need to specify how this action embeds in $G=\UU(N)$. This is equivalent to specify how the $N$-dimensional subspace $W$ in \eqref{Xn} transforms under $\bZ_p \times G$: \begin{equation}
W= e^{2\pi \ii  t_1/p} e^{\ii \beta a_1}+ \cdots+ e^{2\pi \ii  t_N/p} e^{\ii \beta a_N}.\label{ww}
\end{equation}

The moduli space $M_{G,n}$ has a natural action of $\UU(1)^2\times G$, to which we now have an embedding of $\bZ_p$ via \eqref{gg} and \eqref{ww}. Then the moduli space of instantons on the orbifold, $M_{G,n}^g$, is just the $\bZ_p$ invariant part of $M_{G,n}$. 

A fixed point of $M_{G,n}^g$ under $\UU(1)^{2+r}$ is still a fixed point in $M_{G,n}$. Therefore, it is still specified by $W$ and $V$ as in \eqref{WV}. The vector space $V$ now has an action of $g$, which is fixed to be 
\begin{equation}
V_p = \sum_{v=1}^N \sum_{(i,j)\in Y_v} 
e^{2\pi \ii  (t_v+(1-i)s_1+(1-j)s_2) /p}
e^{\ii \beta (a_v +(1-i)\epsilon_1+(1-j)\epsilon_2)}.
\end{equation}
Then, the tangent space at the fixed point and/or the hypermultiplet zero modes can be just obtained by projecting down \eqref{tangent}, \eqref{fa} and \eqref{bif} to the part invariant under the $\bZ_p$ action. 

It is now a combinatorial exercise to write down a general formula for the instanton partition function on the orbifold; as reviewed in the previous section, this includes the case with surface operator. It is again to be said that, however, it is easier to implement the algorithm as written above, than to first write down a combinatorial formula and then implement it in a computer algebra system.

Let us now focus on the case when $(s_1,s_2)=(1,-1)$. Then the orbifold $\bC^2/\bZ_p$ is hyperk\"ahler. Let us consider $\UU(N)$ gauge theory on it.
 We can construct the infinite dimensional space $\cV_{G,p}$ as before, by taking the direct sum of the equivariant cohomology of the moduli spaces of $\UU(N)$ instantons on it.
The vector space $\cV_{G,p}$
  is long known to have an action of the affine algebra $\SU(p)_N$ \cite{Nakajima1,Nakajima2}, but this affine algebra is not enough to generate all the states in $\cV_{G,p}$. It is now believed \cite{Nishioka:2011jk,Belavin:2011sw,Belavin:2012eg} that $\cV_{G,p}$ 
  is a representation of a free boson, $\SU(p)_N$, and the $p$-th para-$W_N$ algebra:
\begin{equation}
\frac{\hat{\SU}(N)_p\times \hat{\SU}(N)_k}{\hat{\SU}(N)_{p+k}}
\end{equation} where $k$ is a parameter determined by the ratio $\epsilon_1/\epsilon_2$.
For $p=2$ and $N=2$, the 2nd para-$W_2$ algebra is the standard $\cN=1$ super Virasoro algebra, and many checks have been made \cite{Belavin:2011pp,Belavin:2011tb,Bonelli:2011jx,Bonelli:2011kv,Belavin:2012aa}.  See also \cite{Pedrini:2014yoa} for the analysis of the case $N=1$ for general $p$. 

\subsection{On non-compact toric spaces}\label{toric}
There is another way to study $G=\UU(N)$ instantons on the $\bZ_p$ orbifolds \eqref{gg}, as they can be resolved to give a smooth non-compact toric spaces $X$, where instanton counting can be performed  \cite{Nakajima:2003pg,Nekrasov,Gasparim:2008ri}. 

The basic idea is to realize that the fixed points under $\UU(1)^{2+N}$ of the  $n$-instanton moduli space $M_{G,n}$ on $\bC^2$ correspond to point-like $n$ instantons at the origin of $\bC^2$, which are put on top of each other.  The deformation of the instanton moduli space was done to deal with this singular configuration in a reliable way.  
The toric space $X$ has an action of $\UU(1)^2$, whose fixed points $P_1$, \ldots, $P_k$ are isolated. The action of $\UU(1)^2$ at each of the fixed points can be different: \begin{equation}
TX|_{P_i} = e^{\ii \beta \epsilon_{1;i}} + e^{\ii \beta \epsilon_{2;i}}
\end{equation} where $\epsilon_{1,2;i}$ are integral linear combinations of $\epsilon_{1,2}$. 
Then an $\UU(N)$ instanton configuration on $X$ fixed under $\UU(1)^{2+N}$, is basically given by 
assigning a  $\UU(N)$-instanton configuration on $\bC^2$, at each $P_i$.  Another  data are the magnetic fluxes  $\vec m_j=(m_{j,1},\ldots,m_{j,N})$ through compact 2-cycles $C_j$ of $X$.
Here it is interesting not just to compute the partition function but also correlation functions of certain operators $\mu(C_j)$ which are supported on $C_j$.
Then the correlation function has a schematic form \begin{equation}
Z_X(\mu(C_1)^{d_1}\mu(C_2)^{d_2}\cdots;\epsilon_1,\epsilon_2) = \sum_{\vec m} q^{\vec m_i C^{ij} \vec m_j} f_{d_1,d_2,\ldots}(\vec m_1,\vec m_2,\ldots) \prod_{P_i}  Z_{\bC^2}(\epsilon_{1;i},\epsilon_{2;i})
\end{equation} where $C^{ij}$ is the intersection form of the cycles $C_j$ and $f_{d_1,d_2,\ldots}(\vec m_1,\vec m_2,\ldots)$ is  a prefactor  expressible in a closed form.  For details, see the papers referred to above. 

\paragraph{Nakajima-Yoshioka}
When $X$ is the blow-up $\hat \bC^2$ of $\bC^2$ at the origin, there are two fixed points $P_1$ and $P_2$, with \begin{equation}
T\hat \bC^2|_{P_1}=e^{\ii \beta\epsilon_1}+e^{\ii \beta(\epsilon_2-\epsilon_1)},\quad
T\hat \bC^2|_{P_2}=e^{\ii \beta(\epsilon_1-\epsilon_2)}+e^{\ii \beta\epsilon_2}.
\end{equation} We have one compact 2-cycle $C$. Then we have a schematic relation \begin{equation}
Z_{\hat\bC^2}(\mu(C)^d;\epsilon_1,\epsilon_2) = \sum_{\vec m}  q^{\vec m\cdot \vec m} f_d(\vec m)
Z_{\bC^2}(\epsilon_1,\epsilon_2-\epsilon_1)
Z_{\bC^2}(\epsilon_1-\epsilon_2,\epsilon_2).\label{KKK}
\end{equation} We can use another knowledge here that the instanton moduli space on $\hat\bC^2$ and that on $\bC^2$ can be related via the map $\hat\bC^2\to \bC^2$. Let us assume that $c_1$ of the bundle on $\hat\bC^2$ is zero. Then we have a relation schematically of the  form \begin{equation}
Z_{\hat\bC^2}(\mu(C)^d;\epsilon_1,\epsilon_2) =
\begin{cases}
 Z_{\bC^2}(\epsilon_1,\epsilon_2) & (d=0),\\
0 & (d>0)
\end{cases}
\label{LLL}
\end{equation} for $d=0,1,\ldots,2N-1$.  The combination of \eqref{KKK} and \eqref{LLL} allows us to  write down a recursion relation of the form \begin{equation}
Z_{\bC^2}(\epsilon_1,\epsilon_2) = \sum_{\vec m} q^{\vec m \cdot \vec m} c({\vec m}) 
Z_{\bC^2}(\epsilon_1,\epsilon_2-\epsilon_1)
Z_{\bC^2}(\epsilon_1-\epsilon_2,\epsilon_2).
\end{equation} This allows one to compute the instanton partition function on $\bC^2$ recursively as an expansion in $q$ \cite{Nakajima:2003pg}, starting from the trivial fact that the zero-instanton moduli space is just a point.  From this, a recursive formula for the prepotential can be found, which was studied and written down in \cite{Losev:1997tp,Lossev:1997bz}. The recursive formula was proved from the analysis of the Seiberg-Witten curve in \cite{Losev:1997tp,Lossev:1997bz}, while it was derived from the analysis of the instanton moduli space  in \cite{Nakajima:2003pg}. This gives one proof that the Seiberg-Witten prepotential as defined by the Seiberg-Witten curve is the same as the one defined via the instanton counting.  This method has been extended to the case with matter hypermultiplets in the fundamental representation \cite{Gottsche:2010ig}.

The recursive formula, although mathematically rigorously proved only for $\SU$ gauge groups, has a form transparently given in terms of the roots of the gauge group involved. This conjectural version of the formula for general gauge groups can then be used to determine the instanton partition function for any gauge group. This was applied to $E_{6,7}$ gauge theories in \cite{Keller:2012da} and the function thus obtained agreed with the one computed via the methods of Sec.~\ref{exceptional}. 

The CFT interpretation of this formula was explored in \cite{Wyllard:2011mn}. A similar formula can be formulated for the orbifolds of $\bC^2$ and was studied in \cite{Bonelli:2012ny}. It is also found that the instanton counting on $\bC^2/\bZ_p$ and that on its blowup can have a subtle but controllable difference \cite{Ito:2013kpa}.

\section*{Acknowledgment}
The author thanks his advisor Tohru Eguchi, for suggesting him to review Nekrasov's seminal work \cite{Nekrasov:2002qd} as a project for his master's thesis.  The interest in instanton counting never left him since then. The Mathematica code to count the instantons which he wrote during that project  turned out to be crucial when he started a collaboration five years later with Fernando Alday and Davide Gaiotto, leading to the paper \cite{Alday:2009aq}. He thanks Hiraku Nakajima for painstakingly explaining  basic mathematical facts to him. He would also like to thank all of his collaborators on this interesting arena of instantons and W-algebras. 

It is a pleasure for the author also to thank Miranda Cheng, Hiroaki Kanno, Vasily Pestun, Jaewon Song, Masato Taki,  Joerg Teschner and Xinyu Zhang for carefully reading an early version of this review and for giving him constructive and helpful comments.

This work  is supported  in part by JSPS Grant-in-Aid for Scientific Research No. 25870159 and in part by WPI Initiative, MEXT, Japan at IPMU, the University of Tokyo.

\bibliographystyle{ytphys}
\small\baselineskip=.93\baselineskip
\let\bbb\bibitem\def\bibitem{\itemsep1pt\bbb}

\paragraph{\large References to articles in this volume}
\renewcommand{\refname}{

\paragraph{Other references}
\renewcommand{\refname}{\vskip-36pt}
\bibliography{ref}

\end{document}